\documentclass{article}
\pdfoutput=1
\usepackage[T1]{fontenc}
\usepackage[utf8]{inputenc}
\usepackage{bm}
\usepackage{amsmath}
\usepackage{amssymb}
\usepackage[english]{babel}
\usepackage{graphicx}
\usepackage{subscript}
\usepackage{color,soul}
\usepackage{booktabs}
\usepackage{caption}
\usepackage{subcaption}
\usepackage{authblk}
\usepackage{algorithm} 
\usepackage{algpseudocode}

\usepackage[backend=bibtex, natbib=true, bibstyle=numeric, citestyle=numeric, maxbibnames=99, doi=true, sorting=none]{biblatex}
\begin{document}
\selectlanguage{english}

\title{Referenced Thermodynamic Integration for Bayesian Model Selection: Application to COVID-19 Model Selection}

 \author{Iwona Hawryluk$^*$, Swapnil Mishra, Seth Flaxman, Samir Bhatt$^{\dagger}$, and Thomas A. Mellan$^{\dagger *}$}
\affil{MRC Centre for Global Infectious Disease Analysis, Department of Infectious Disease Epidemiology, Imperial College London}
\affil{Department of Mathematics, Imperial College London}
\affil{$^\dagger$ Correspondence: t.mellan@imperial.ac.uk, \\s.bhatt@imperial.ac.uk}
\affil{$^*$ Contributed equally}

\date{}

\maketitle

\begin{abstract}

Model selection is a fundamental part of the applied Bayesian statistical methodology. Metrics such as the Akaike Information Criterion are commonly used in practice to select models but do not incorporate the uncertainty of the models’ parameters and can give misleading choices. One approach that uses the full posterior distribution is to compute the ratio of two models’ normalising constants, known as the Bayes factor. Often in realistic problems, this involves the integration of analytically intractable, high-dimensional distributions, and therefore requires the use of stochastic methods such as thermodynamic integration (TI). In this paper we apply a variation of the TI method, referred to as \emph{referenced TI}, which computes a single model's normalising constant in an efficient way by using a judiciously chosen reference density. The advantages of the approach and theoretical considerations are set out, along with explicit pedagogical 1 and 2D examples. Benchmarking is presented with comparable methods and we find favourable convergence performance. The approach is shown to be useful in practice when applied to a real problem — to perform model selection for a semi-mechanistic hierarchical Bayesian model of COVID-19 transmission in South Korea involving the integration of a 200D density. \\
\end{abstract}

\newpage

\section{Introduction}
\let\thefootnote\relax\footnote{Code available at \url{https://github.com/mrc-ide/referenced-TI}}
Bayesian computation has been in the limelight in recent months. Modern techniques of statistical machine learning, in particular those based on stochastic inference, have been central to current epidemiological models \citep{sharma2020robust, flaxman2020estimating}, with statistical inference underpinning estimates of pathogen transmission being used to make informed public health policy.
With advances in both computing power and availability of data, it is possible to build more complex, reliable and accurate models, while the recent increased focus on epidemiological models emphasises a need for synthesis with rigorous statistical methods. This synthesis is required to robustly estimate necessary parameters, quantify uncertainty in predictions, and to test hypotheses \citep{grassly_review}. 

In practice the decision to choose a model is often based on heuristics, relying on the knowledge and experience of the modeller, rather than through a systematic selection process \citep{sun_wildlife, rotavirus_germany}. A number of model selection methods are available, but those methods often come with a trade-off between accuracy and computational complexity. For example, widely used in epidemiology Akaike Information Criterion (AIC) or Bayesian Information Criterion (BIC) are easy to compute, but come with certain limitations \citep{grassly_review}. Specifically, they do not take into account the parameters' uncertainty or the prior probabilities, and might favour excessively complex models.

The ratio of two normalising constants --- the Bayes factor (BF) --- is a popular and general method used for model selection in the Bayesian setting \citep{bayes_factors}. This approach integrates out the parameters of the model to find the probability of the getting the data from a given model hypothesis. But in general a normalising constant cannot be computed analytically or through direct quadrature methods, due to the difficulty of integrating arbitrary high-dimensional distributions. Typically approximations are applied such as Laplace's method, variational approximations, or probabilistic algorithms such as bridge sampling \citep{bennett1976efficient,meng1996simulating}, stochastic density of states based methods \citep{skilling_nested_sampling, habeck2012evaluation} or thermodynamic integration \citep{TI_orig,gelman1998simulating,friel_improving, lartillot}.

Thermodynamic integration (TI) provides a useful way estimate the
log ratio of the normalising constants of two densities. Instead of marginalising the densities explicitly in terms of the high-dimensional integrals, by using TI we only need to evaluate a 1-dimensional integral, where the integrand can easily be sampled with Markov Chain Monte Carlo (MCMC). To see
how this works, consider two models with the pair of normalising constants $z_{1}$
and $z_{2}$, 
\begin{equation}
z_{i}=\int q_{i}(\bm{\theta})d\bm{\theta}\,,\,\,i\in\{1,2\}\,,\label{eq: z_i_definition}
\end{equation}
where $q_{i}$ is a density for model $M_i$ with parameters $\bm{\theta}$, that can be normalised to give the model's Bayesian posterior density
\[
p_{i}(\bm{\theta})=\frac{q_{i}(\bm{\theta})}{z_{i}}\,,\,\,i\in\{1,2\}\,.
\]
To apply thermodynamic integration we introduce the concept of a path between $q_{1}(\bm{\theta})$ and $q_{2}(\bm{\theta})$, linking the two densities via a series of intermediate ones. This family of densities is denoted $q(\lambda;\bm{\theta})$. An example path in $\lambda$ is shown in Figure \ref{fig: 1dcusp}. Note --- we use the symbol $\lambda$ but often the coupling parameter is denoted $\beta$ (or $t$) in reference to a physical thermodynamic integration in inverse temperature (or temperature). In learning terms the temperature represents a Lagrange multiplier regularising the loss, and in many instances the tempering analogy is useful. Here however the approach is a more generic procedure and we prefer to consider it simply as a path integration between two distributions coupled by an arbitrary but common switching parameter $\lambda$.

The parametric density $q(\lambda;\bm{\theta})$, linking
$q_{1}$ to $q_{2}$ and defining the intermediate densities, can
be chosen to have an optimal or in some way convenient path, but a
common choice is the geometric one
\[
q(\lambda;\bm{\theta})=q_{2}^{\lambda}(\bm{\theta})q_{1}^{1-\lambda}(\bm{\theta})\,,\,\,\lambda\in[0,1]\,.
\]
The important point to note is that for $\lambda=0$, $q(\lambda;\bm{\theta})$ returns the first density $q(0;\bm{\theta})=q_{1}(\bm{\theta})$,
for $\lambda=1$ it gives $q(1;\bm{\theta})=q_{2}(\bm{\theta})$,
and for in-between $\lambda$ values a log-linear mixture of the
endpoint densities. Just as we have defined a family of densities, there is an associated normalising constant for any point along the
path, that for any value of $\lambda$ is given by 
\[
z(\lambda)=\int_{\Omega(\lambda)} q(\lambda;\bm{\theta})d\bm{\theta}\,.
\]
A further small but important point to avoid complications is to have densities that have common support, for example ${\Omega(1)} = {\Omega(0)}$. Hereafter support is denoted $\Omega$.

Having set up the definitions of $q(\lambda;\bm{\theta})$ and $z(\lambda)$, the TI expression can be derived, to compute the
log ratio of $z_1=z(0)$ and $z_2=z(1)$, while avoiding explicit integrals over the models' parameters $\bm{\theta}$. By the fundamental theorem
of calculus, assuming that the order of $\partial_{\lambda}$  and
$\int d\bm{\theta}$ integral can be exchanged, and by the rules
of differentiating logarithms we get:
\begin{eqnarray}
\text{log}\,\frac{z_{2}}{z_{1}} & = & \int_{0}^{1}\partial_{\lambda}\text{log\,}z(\lambda)\,d\lambda\nonumber \\
 & = & \int_{0}^{1}\frac{1}{z(\lambda)}\partial_{\lambda}z(\lambda)\,d\lambda\nonumber \\
 & = & \int_{0}^{1}\frac{1}{z(\lambda)}\partial_{\lambda}\int_{\Omega}q(\lambda;\bm{\theta})\,d\bm{\theta}\,d\lambda\nonumber \\
 & = & \int_{0}^{1}\frac{1}{z(\lambda)}\int_{\Omega}\left(\partial_{\lambda}\text{log}\,q(\lambda;\bm{\theta})\right)q(\lambda;\bm{\theta})\,d\bm{\theta}\,d\lambda\nonumber \\
 & = & \int_{0}^{1}\mathbb{E}_{ p(\lambda;\bm{\theta})}\left[\partial_{\lambda}\text{log\,}q(\lambda;\bm{\theta})\right]\,d\lambda\nonumber \\
 & = & \int_{0}^{1}\mathbb{E}_{ p(\lambda;\bm{\theta})}\left[\text{log\,}\frac{q_{2}(\bm{\theta})}{q_{1}(\bm{\theta})}\right]\,d\lambda\nonumber \\
 & = & \int_{0}^{1}\mathbb{E}_{ q(\lambda;\bm{\theta})}\left[\text{log\,}\frac{q_{2}(\bm{\theta})}{q_{1}(\bm{\theta})}\right]\,d\lambda\,.\label{eq: TI_derivation}
\end{eqnarray}
The notation $\mathbb{E}_{q(\lambda;\theta)}$ is for an expectation with respect to the sampling distribution $q(\lambda; \theta)$. The final line in the expression summarises the usefulness of TI---instead of having to work with the complicated high-dimensional
integrals of Equation \ref{eq: z_i_definition} to find the log-Bayes
factor $\text{log}\,\frac{z_{2}}{z_{1}}$, we only need consider
a 1-dimensional integral of an expectation, and the expectation can
be readily produced by MCMC.

Here we set out in detail some variations on the TI theme that we have found to be useful in practice, in particular with regard to evaluating evidence for epidemiological models. The variation is to work primarily in terms
of appropriately chosen reference normalising constants. The approach of using exactly integrate-able references has provided us with a particularly
efficient method of selection between different hierarchical Bayesian models models, and we hope the approach will be useful to others working on similar problems, for example in phylogenetic model selection where TI is already a popular established method \citep{phylo_bayes, lartillot}. 

In the following, we begin with introducing the reference density, then go on to illustrate different practical choices of a reference normalising constant, along with theoretical and practical considerations. Next, the mechanics of applying the method
are set out for two simple pedagogical examples.
Performance benchmarks are discussed for a well-known problem in the statistical literature \citep{radiata}, which shows the method performs favourably in terms of accuracy and the number of iterations to convergence. Finally the technique
is applied to a hierarchical Bayesian time-series model describing
the COVID-19 epidemic in South Korea. In the COVID-19 infections model
we show how the approach may be used to select technical parameters that specify the reproduction number ($R_t$) model, such as the autoregression window size and AR($k$) lag, as well
epidemiologically meaningful parameters such as the serial
interval distribution time for generating infections.

\section{Referenced-TI}

An efficient approach to compute Bayes factors, or more generally to marginalise a given density for any application, is to introduce a reference as 
\begin{eqnarray}
z & = & z_{\text{ref}}\,\frac{z}{z_{\text{ref}}}\nonumber \\
 & = & z_{\text{ref}}\,\text{exp}\int_{0}^{1}\mathbb{E}_{ q(\lambda;\bm{\theta})}\left[\text{log\,}\frac{q(\bm{\theta})}{q_{\text{ref}}(\bm{\theta})}\right]\,d\lambda\,.\label{eq: reference_TI}
\end{eqnarray}
To clarify notation, $z$ is the normalising constant of interest with density $q$,
$z_{\text{ref}}$ is a reference normalising constant with associated density $q_{\text{ref}}$. The second line replaces the ratio $z/z_\text{ref}$ with a thermodynamic integral as per the identity derived in Equation \ref{eq: TI_derivation}.

The introduction of a reference naturally facilitates the marginalisation
of a single density, rather than requiring pairwise model comparisons
by direct application of Equation \ref{eq: TI_derivation}. This is useful when comparing multiple models as $n>\binom{n}{2}$ for $n\ge3$.
Another motivation to reference the TI is the MCMC computational efficiency of converging the TI expectation. In Equation \ref{eq: reference_TI},
with judicious choice of $q_{\text{ref}}$, the reference normalising constant
$z_{\text{ref}}$ can be evaluated analytically and account for most
of $z$. In this case $\text{log\,}\frac{q(\bm{\theta})}{q_{\text{ref}}(\bm{\theta})}$
tends to have a small expectation and variance and converges quickly.

The idea of using an exactly solvable reference, to aid in the solution of an otherwise intractable problem, is a recurrent and perennial theme in the computational and mathematical sciences in general \citep{dirac1927quantum,gell1957correlation,vocadlo2002ab,duff2015improved}, and variations on this approach have been used to compute normalising constants in various guises in the statistical literature \citep{power_posteriors_original,friel_improving,friel_review,lartillot,phylo_bayes, cameron2014recursive, lefebvre2010path, GSS_fan_2011, GSS_baele}. For example, in the generalised stepping stone method a reference is introduced to speed up convergence of the importance sampling at each temperature rung \citep{GSS_fan_2011, GSS_baele}. In the work of Lefebvre et al. \citep{lefebvre2010path} a theoretical discussion has been presented  that shows the error budget of thermodynamic integration depends on the J-divergence of the densities being marginalised. Noting this, Cameron and Pettitt \citep{cameron2014recursive} illustrate in their work on recursive pathways to marginal likelihood estimation, that a "data driven" auxiliary density reduces standard errors for a 2D banana-shaped likelihood posterior. While, in the power posteriors (PP) approach, the reference in Equation \ref{eq: reference_TI} is the prior distribution of $q$ and thus $z_{\text{ref}}=1$ \citep{power_posteriors_original}. This approach is elegant as the reference need not be chosen --- it is simply the prior --- however the downside of the simplicity is that for poorly chosen or uninformative priors, the thermodynamic integral will be slow to converge and susceptible to instability. In particular for complex hierarchical models with weakly informative priors this is found to be an issue.

The reference density in Equation \ref{eq: reference_TI} can be chosen
at convenience, but the main desirable features are that it should
be easily formed without special consideration or adjustments, similar
to the power posteriors method, and that $z_{\text{ref}}$ should be analytically integratable and account for as much of $z$ as possible. Such a choice of $z_{\text{ref}}$ ensures the part with expensive sampling is small and convergence is fast. An obvious choice in this regard is the Laplace-type reference, where the log density is approximated
with a second-order one, for example a multivariate Gaussian. For densities
with a single concentration, Laplace-type approximations are ubiquitous, and an excellent natural choice for many problems. In the following section we consider three approaches
that can be used to formulate a reference normalising constant $z_{\text{ref}}$
from a second-order log density (though more generally other tractable references are possible). In each referenced TI scenario, we
note that even if the reference approximation is poor, the estimate of
the normalising constant based on Equation \ref{eq: reference_TI}
remains asymptotically exact---only the speed of convergence may
be reduced (subject to the assumptions of matching support of endpoint
densities).

\subsection{Taylor Expansion at the Mode Laplace Reference}

The most straightforward way to generate a reference density is to
Taylor expand $\text{log}\,q(\bm{\theta})$ to second order about a mode. Noting there is no linear term, we see the reference
density is
\begin{eqnarray}
q_{\text{ref}}(\bm{\theta}) & = & \text{exp}\left(\text{log}\,q(\bm{\theta}_{0})+\frac{1}{2}(\bm{\theta}-\bm{\theta}_{0})^{\text{T}}\mathbf{H}\,(\bm{\theta}-\bm{\theta}_{0})\right)\,,
\end{eqnarray}
where $\mathbf{H}$ is the hessian matrix and $\bm{\theta}_{0}$ is
the vector of mode parameters. The associated normalising constant
is
\begin{eqnarray}
z_{\text{ref}} & = & \int_{-\infty}^{\infty}q_{\text{ref}}(\bm{\theta})d\bm{\theta}\nonumber\,\\
 & = & \text{\ensuremath{\int_{-\infty}^{\infty}}exp}\left(\text{log}\,q(\bm{\theta}_{0})+\frac{1}{2}\left(\bm{\theta}-\bm{\theta}_{0}\right)^{\text{T}}\mathbf{H}\left(\mathbf{\bm{\theta}}-\bm{\theta}_{0}\right)\right)d\bm{\theta}\nonumber\\
 & = & \text{q(\ensuremath{\bm{\theta}_{0}})\ensuremath{\int_{-\infty}^{\infty}}exp}\left(\frac{1}{2}\left(\bm{\theta}-\bm{\theta}_{0}\right)^{\text{T}}\mathbf{H}\left(\mathbf{\bm{\theta}}-\bm{\theta}_{0}\right)\right)d\bm{\theta}\nonumber\\
 & = & q(\bm{\theta}_{0})\sqrt{\text{det}(2\pi\mathbf{H}^{-1})}\,.\label{eq: reference_integrated}
\end{eqnarray}
The Taylor expansion method at the mode with analytic or finite difference gradients tends to produce a reference density that
works well in the neighbourhood of $\bm{\theta}_{0}$ but can be less
suitable if density is asymmetric, has long or short tails, or if the derivatives
at the mode are poorly approximated for example due to cusps or conversely
very flat curvature at the mode. In many instances a more reliable choice
of reference can found by using MCMC samples from the whole posterior
density.

\subsection{Sampled Covariance Laplace Reference}

The second straightforward but often more robust approach is to form a reference density by
drawing samples from the true density $q(\bm{\theta})$, to estimate the mean parameters $\hat{\bm{\theta}}$ and covariance matrix
$\hat{\bm{\Sigma}}$, such that
\begin{eqnarray}
q_{\text{ref}}(\bm{\theta}) & = & q(\hat{\bm{\theta}})\,\text{exp}\left(-\frac{1}{2}(\bm{\theta}-\hat{\bm{\theta}})^{\text{T}}\bm{\hat{\Sigma}}^{-1}\,(\bm{\theta}-\hat{\bm{\theta}})\right)\,.\label{eq: covariance_laplace}
\end{eqnarray}
The reference normalising constant is $z_{\text{ref}}=q(\hat{\bm{\theta}})\sqrt{\text{det}(2\pi\hat{\bm{\Sigma}})}\,$.

This method of generating a reference is simple and reliable. It
requires sampling from the posterior $q(\bm{\theta})$ so is more expensive
than derivative methods, but the cost associated with drawing enough
samples to generate a sufficiently good reference tends to be quite
low. In the primary application discussed later, regarding relatively
complex high-dimensional Bayesian hierarchical models, we use this
approach to generate a reference density and normalising constant. 

The sampled covariance reference is typically a good approach,
but it is not in general optimal within the Laplace-type family of approaches --- typically another Gaussian reference exists with different parameters that can generate a normalising
constant closer to the true one, thus potentially leading to overall faster convergence
of the thermodynamic integral to the exact value. Such an optimal
reference can be identified variationally. 

\subsection{Variational Laplace Reference}

The conditions to identify an optimal reference normalising constant can be derived by considering a Taylor expansion
of the log normalising constant $\text{log}\,z(\lambda)$
about $\lambda=0$:
\[
\text{log}\,z(\lambda)\approx\text{log}\,z(0)+\lambda\,\partial_{\lambda}\text{log}\,z(0)+\frac{1}{2}\lambda^{2}\,\partial_{\lambda}^{2}\text{log}\,z(0)\,.
\]
The first derivative gives the expectation
\[
\text{\ensuremath{\partial_{\lambda}}log\,}z(\lambda) = \mathbb{E}_{ q(\lambda;\bm{\theta})}\left[\text{log\,}\frac{q(\bm{\theta})}{q_{\text{ref}}(\bm{\theta})}\right]\,,
\]
as per the derivation in Equation \ref{eq: TI_derivation}, and the second derivative is a variance
\begin{eqnarray*}
\text{\ensuremath{\partial_{\lambda}^{2}}log\,}z(\lambda) & = & \frac{\int\left(\text{log}\,\frac{q(\bm{\theta})}{q_{\text{ref}}(\bm{\theta})}\right)^{2}q(\lambda;\bm{\theta})d\bm{\theta}}{\int q(\lambda;\bm{\theta})d\bm{\theta}}-\left(\frac{\int\left(\text{log}\,\frac{q(\bm{\theta})}{q_{\text{ref}}(\bm{\theta})}\right)\,q(\lambda;\bm{\theta})d\bm{\theta}}{\int q(\lambda;\bm{\theta})d\bm{\theta}}\right)^{2}\\
 & = & \left\{ \mathbb{E}_{ q(\lambda;\bm{\theta})}\left[\left(\text{log}\,\frac{q(\bm{\theta})}{q_{\text{ref}}(\bm{\theta})}\right)^{2}\right]-\mathbb{E}_{ q(\lambda;\bm{\theta})}\left[\text{log}\,\frac{q(\bm{\theta})}{q_{\text{ref}}(\bm{\theta})}\right]^{2}\right\} \\
 & \ge & 0\,.
\end{eqnarray*}
As the curvature
of $\text{log}\,z(\lambda)$ is increasing, to first order
we see
\[
\text{log}\,z(\lambda)\ge\text{log\,\ensuremath{z(0)}+\ensuremath{\lambda\mathbb{E}_{ q(0;\bm{\theta})}\left[\text{log}\,\frac{q(\bm{\theta})}{q_{0}(\bm{\theta})}\right]}}\,,
\]
and for the specific case of $\lambda=1$,
\[
\text{log}\,z\ge\text{log\,\ensuremath{z_{\text{ref}}}+\ensuremath{\mathbb{E}_{ q_{\text{ref}}(\bm{\theta})}\left[\text{log}\,\frac{q(\bm{\theta})}{q_{\text{ref}}(\bm{\theta})}\right]}}\,.
\]
This inequality establishes bounds that can be maximised with respect
to the position ($\bm{\mu}$) and scale ($\bm{S}$) parameters of
a reference density such as
\[
q_{\text{ref}}(\bm{\mu},\bm{S};\bm{\theta})=q(\bm{\mu})\text{exp}\left(-\frac{1}{2}(\bm{\theta}-\bm{\mu})^{\text{T}}\bm{S}\,(\bm{\theta}-\bm{\mu})\right)\,.
\]
Thus the parameters that optimise
\[
\underset{\bm{\mu},\bm{S}}{\text{max}}\left\{ \text{log}\,z_{\text{ref}}+\mathbb{E}_{q_{\text{ref}}(\bm{\theta})}\left[\text{log}\,\frac{q(\bm{\theta})}{q_{\text{ref}}(\bm{\mu},\bm{S};\bm{\theta})}\right]\right\} \,,
\]
provide a reference density that is variationally optimal. We note this is an application of what is known as the Gibbs-Feynman-Bogoliubov
inequality from other fields \citep{bogolubov1966model,kuzemsky2015variational,zhang1996application}, and that finding approximations of this type to the true density is a well-studied problem in machine learning, with well-documented approaches that can be used to determine $q_{\text{ref}}$ variationally \citep{neal1998view,jordan1999introduction}. In itself the existence of a variational bound provides no guarantee of being a good approximation to the true normalising constant, and is thus alone not a satisfactory general approach. However as a point of reference from which to estimate the true normalising constant, it provides a first-order optimal density within the family of trial reference functions considered, therefore improving convergence to the MCMC normalising constant in referenced TI.

\subsection{Multi-reference TI}

Having set out three approaches to find a single reference for the TI expression in
Equation \ref{eq: reference_TI}, a natural generalisation is the telescopic expansion 
\begin{eqnarray}
z & = & z_{0}\,{\prod_{i=0}^{n-1}}\frac{z_{i+1}}{z_{i}}\frac{z}{z_{n}}\nonumber \\
 & = & z_{0}\,\text{exp}\left(\int_{0}^{1}{\sum_{i=0}^{n-1}}\mathbb{E}_{ q_{i}(\lambda;\bm{\theta})}\left[\text{log\,}\frac{q_{i+1}(\bm{\theta})}{q_{i}(\bm{\theta})}\right]+\mathbb{E}_{ q_{n}(\lambda;\bm{\theta})}\left[\text{log\,}\frac{q(\bm{\theta})}{q_{n}(\bm{\theta})}\right]\,d\lambda\right)\,.\label{eq: reference_TI-telescopic}
\end{eqnarray}
Note, here the analytic reference is denoted $z_{0}$ rather than
$z_{\text{ref}}$ to generalise the indexing. In cases where $q_{0}$ differs
substantially from $q$, the telescopic generalisation can improve
numerical stability. By bridging the endpoints in terms of intermediate
density pairs, $\frac{q_{i+1}(\bm{\theta})}{q_{i}(\bm{\theta})}$,
we can form a series of lower variance MCMC simulations with favourable
convergence properties. A reasonable choice for generating intermediate densities is for the $q_{i^{\text{th}}}$
density to be the $2\left(i+1\right)^{\text{th}}$ order Taylor expansion
of $q(\bm{\theta})$.

\subsection{Reference Support}

If a model has parameters with limits, for example $\theta_{1}\in[0,\infty)$,
$\theta_{2}\in(-1,\infty)$ etc., in referenced TI the exact analytic integration for the reference density
should be commensurately limited. However, the calculation of arbitrary
probability density function orthants, even for well-known analytic
functions such as the multivariate Gaussian, is in general a difficult problem. Computing high-dimensional orthants usually requires advanced techniques, the use of approximations, or sampling methods \citep{ridgway2016computation, azzimonti2018estimating, owen2014orthant,miwa2003evaluation, curnow1962numerical, ruben1964asymptotic}.
Fortunately we can simplify our reference density to create a reference
with tractable analytic integration for limits by using a diagonal approximation to the sampled covariance or hessian matrix. For example the orthant of a diagonal
multivariate Gaussian can be given in terms of the error function \citep{error_function_missile}, leading to the expression
\begin{eqnarray}
z_{\text{ref}}&=&q(\hat{\bm{\theta}})\sqrt{\text{det}(2\pi\bm{\Sigma}^{\text{diag}})}\prod_{i\in K}\left(1+\text{erf}\left(\frac{\hat{\theta}_{i}-a_{i}}{\sqrt{2\Sigma_{i}^{\text{diag}}}}\right)\right)\,,\label{eq: orthant}
\end{eqnarray}
where $K$ denotes the set of indices of the parameters with lower limits $a_{i}$. $\Sigma^{\text{diag}}$ is a diagonal covariance matrix, that is one containing only the variance of each of the parameters, without the covariance terms and $\Sigma_{i}^{\text{diag}}$ denotes the $i^{\text{th}}$ element of the diagonal.
Restricting our density to a diagonal one is a poorer approximation
than using the full covariance matrix. In practice however this has not been particularly
detrimental to the convergence of the thermodynamic integral---and again we
note that the quality of the reference only affects convergence rather
than eventual theoretical Monte Carlo accuracy of the normalising constant. This behaviour is observed in the practical examples later considered, though the distinction between accuracy and convergence and matters of asymptotic consistency using an MCMC estimator with finite iterations are naturally less clear cut.

\subsection{Technical Implementation}
The referenced TI algorithm was implemented in Python and Stan programming languages. Using Stan enables fast MCMC simulations, using Hamiltonian Monte Carlo (HMC) and No-U-Turn samples (NUTS) algorithm \citep{nuts, carpenter2017stan}, and portability between other statistical languages, such as R or Julia. Additionally it is familiar to many epidemiologists using Bayesian statistics \citep{bayes-epi}. The code for all examples shown in this paper is available at \url{https://github.com/mrc-ide/referenced-TI}.
In the examples shown in Section \ref{sec: Application}, we used 4 chains with 20,000 iterations per chain for the pedagogical examples, and 4 chains with 2,000 iterations for the other applications. In all cases, half of iterations were used for the burn-in. Mixing of the MCMC chains and the sampling convergence was checked in each case, by ensuring that the $\hat{R}$ value was $\leq 1.05$ for each parameter in every model.

In all examples in the remaining part of this paper, the integral given in Equation \ref{eq: TI_derivation} was discretised to allow computer simulations. Each expectation $\mathbb{E}_{ q(\lambda;\theta)}\left[\text{log\,}\frac{q_{1}(\theta)}{q_{0}(\theta)}\right]$ was evaluated at $\lambda = 0.0, 0.1, 0.2, .... , 0.9, 1.0$, unless stated otherwise. To obtain the value of the integral in Equation \ref{eq: TI_derivation}, we interpolated a curve linking the expectations using a cubic spline, which was then integrated numerically. The pseudo-code of the algorithm with sampled covariance Laplace reference is shown in Algorithm \ref{algo: algorithm}.

\begin{algorithm}
	\caption{Referenced thermodynamic integration algorithm \label{algo: algorithm}}
	 \hspace*{\algorithmicindent} \textbf{Input} $q$ - un-normalised density, $q_{\text{ref}}$ - un-normalised reference density, $\Lambda$ - set of coupling parameters $\lambda$, $N$ - number of MCMC iterations\\
  \hspace*{\algorithmicindent} \textbf{Output} $z$ - normalising constant of the density $q$
	\begin{algorithmic}[1]
	    \State Define un-normalised density $q$ and the reference density
	    $q_{\text{ref}}$
	    \State Calculate $z_{\text{ref}}$ analytically by using the determinant of the covariance matrix.
		\For {$\lambda \in \Lambda$}
		    \State Sample N values $\theta_n$ from $q^\lambda q_{\text{ref}}^{1-\lambda}$
			\For {$n=1,2,\ldots,N$}
				\State Calculate $\text{log}\frac{q(\theta_n)}{q_{\text{ref}}(\theta_n)}$
			\EndFor
			\State Compute the mean, $\mathbb{E}_\lambda = \frac{1}{N} \Sigma_{n=1}^{N} \text{log}\frac{q(\theta_n)}{q_{\text{ref}}(\theta_n)}$
			\EndFor
		\State Interpolate between the consecutive $\mathbb{E}_\lambda$ values to obtain a curve $\partial_{\lambda} \text{log} (z(\lambda))$
		\State Integrate $\partial_{\lambda} \text{log} (z(\lambda))$ over $\lambda \in [0,1]$ to get $\text{log}\frac{z}{z_{\text{ref}}}$
		\State Calculate $z$ = $z_{\text{ref}} \cdot \text{exp}\{\text{log}\frac{z}{z_{\text{ref}}}\}$
	\end{algorithmic} 
\end{algorithm}

\section{Applications}\label{sec: Application}
In this section we present an application of the referenced TI algorithm to 1- and 2-dimensional pedagogical examples, a linear regression model, and a model selection task for a model of COVID-19 epidemic in South Korea.

\subsection{1D Pedagogical Example}
\begin{figure}
    \begin{subfigure}[t]{0.5\textwidth}
        \centering
        \includegraphics[width=\linewidth]{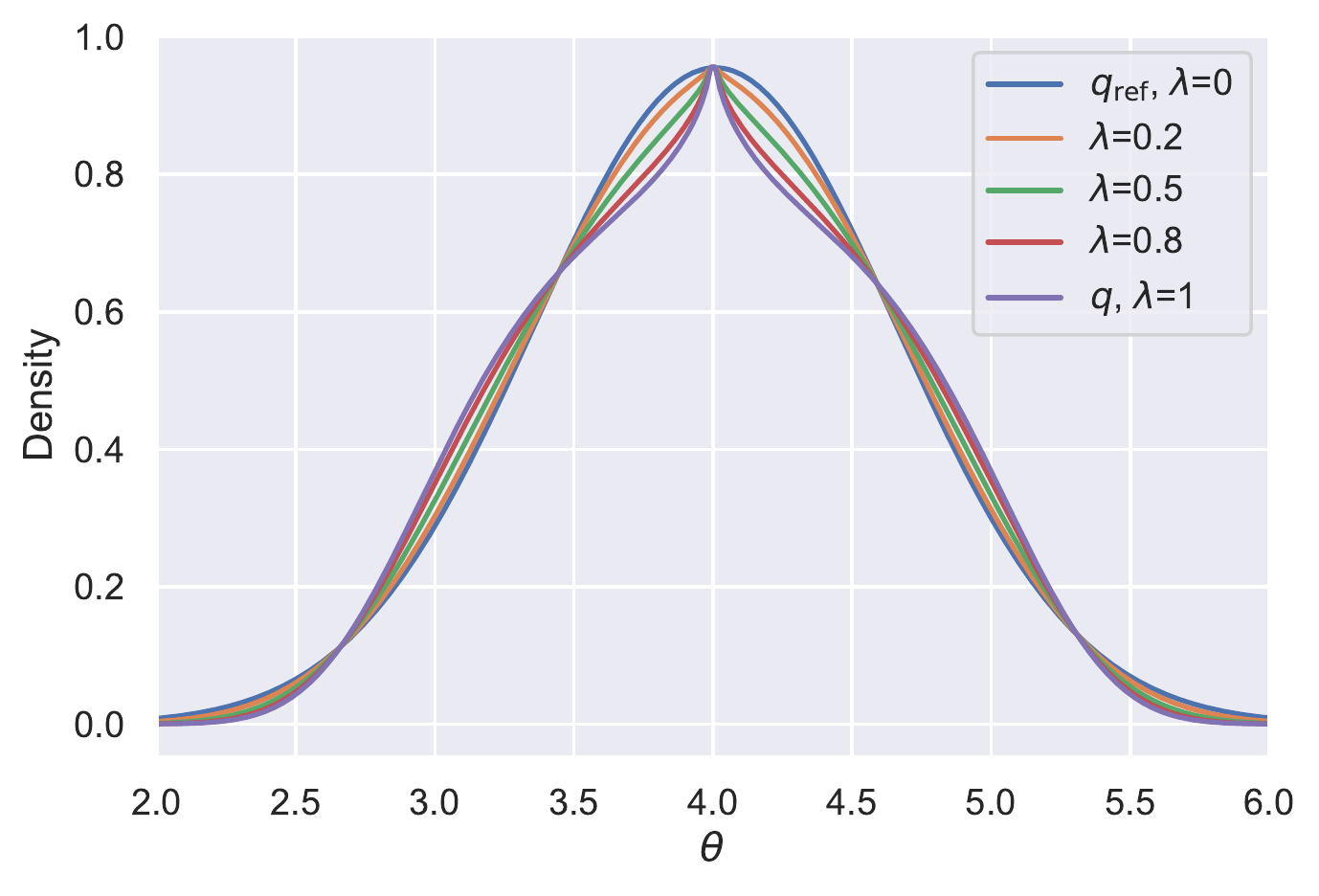}
        \caption{} 
    \end{subfigure}
    \hfill
    \begin{subfigure}[t]{0.5\textwidth}
        \centering
        \includegraphics[width=\linewidth]{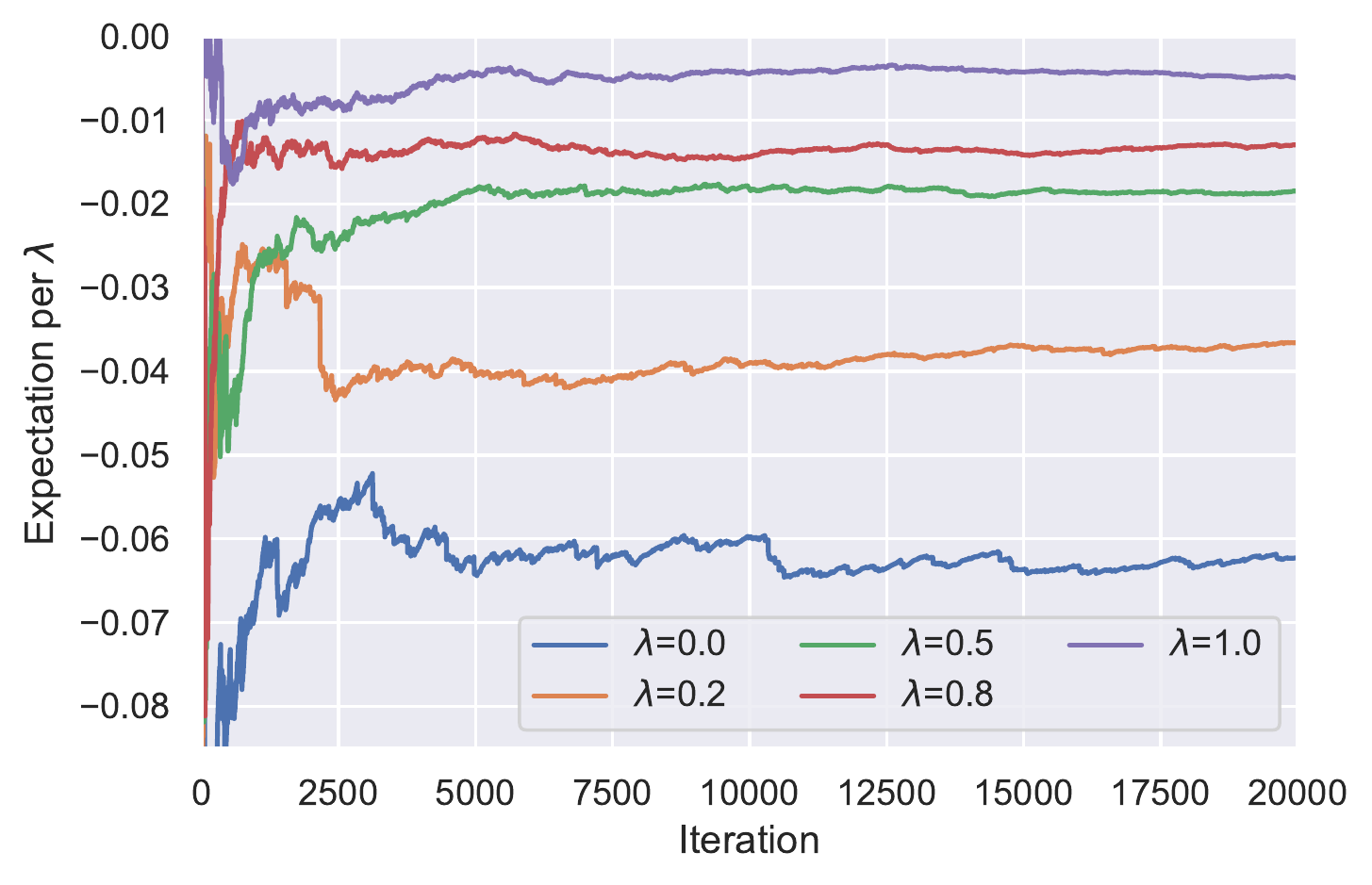} 
        \caption{}
    \end{subfigure}
    \hfill
    \begin{subfigure}[t]{0.5\textwidth}
        \centering
        \includegraphics[width=\linewidth]{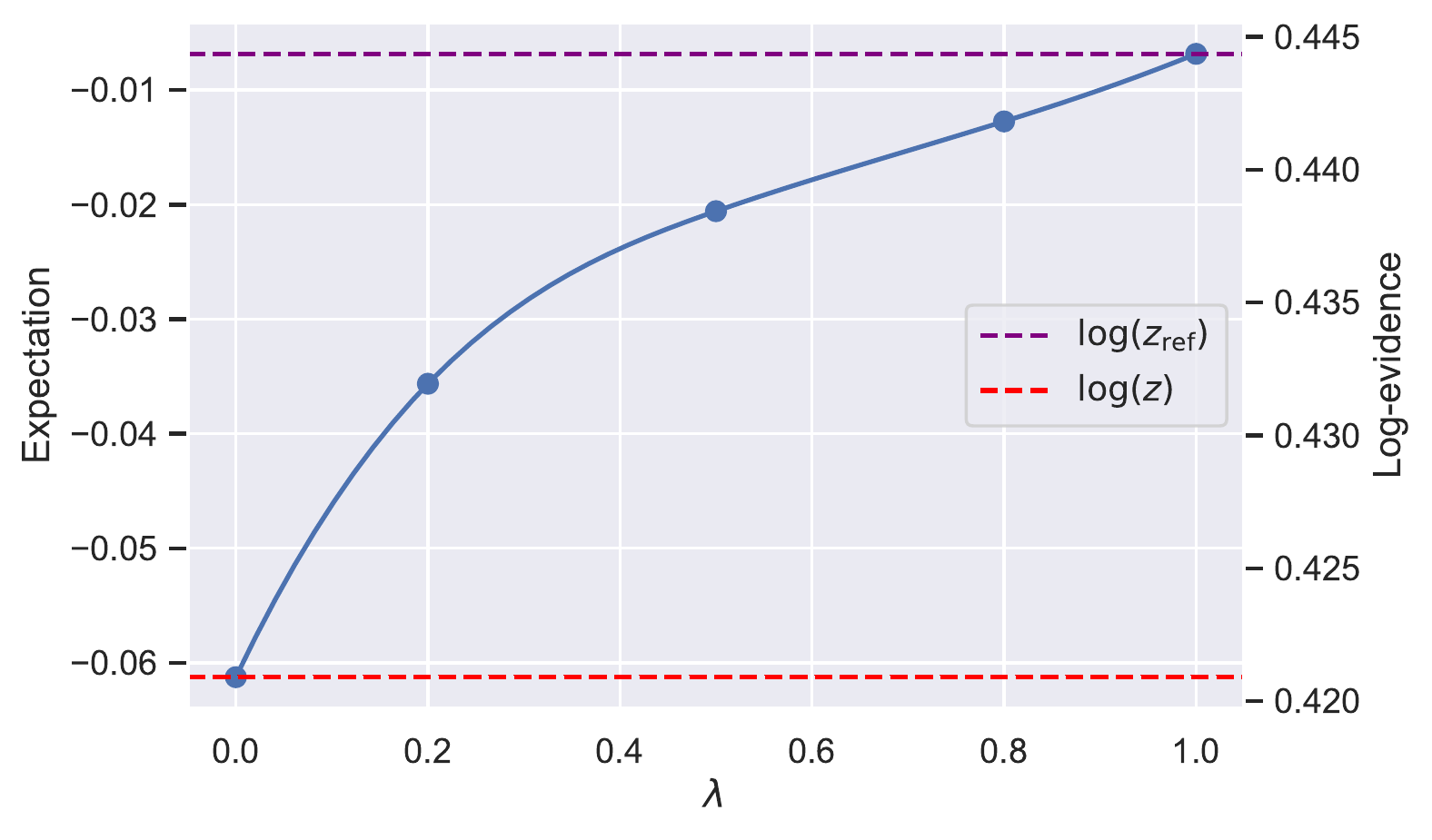} 
        \caption{} 
    \end{subfigure}
    \hfill
    \begin{subfigure}[t]{0.5\textwidth}
        \centering
        \includegraphics[width=\linewidth]{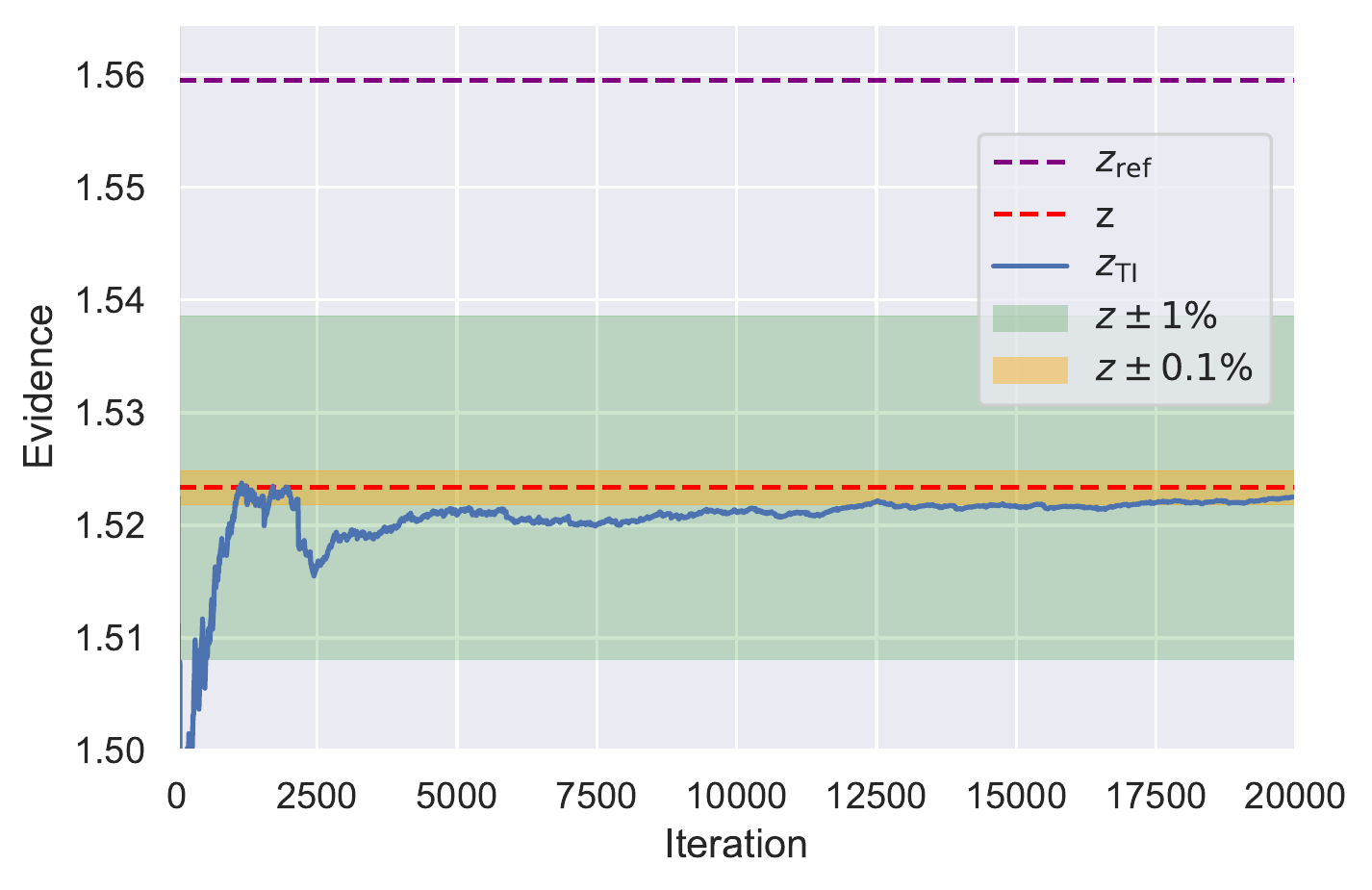} 
        \caption{} 
    \end{subfigure}
    \caption{a) $q^{\lambda}q_{\text{ref}}^{(1-\lambda)}$ for the 1d example density (Equation \ref{eq: 1dexample_density}) at selected $\lambda$ values along the path. b) Expectation $\mathbb{E}_{ q(\lambda;\bm{\theta})}\left[\text{log\,}\frac{q(\bm{\theta})}{q_{\text{ref}}(\bm{\theta})}\right]$ vs MCMC iteration, shown at each value of $\lambda$ sampled. c) $\lambda$-dependence of the TI contribution to the log-evidence d) Convergence of the evidence $z$, with $1$\% convergence after $500$ iterations and $0.1$\% after $17,000$ iterations per $\lambda$.}\label{fig: 1dcusp}
\end{figure}

To illustrate the technique consider the 1-dimensional density
\begin{eqnarray}
q(\theta)=\text{exp}\left(-\frac{1}{2}\sqrt{\left|\theta-4\right|}-\frac{1}{2}\left(\theta-4\right)^{4}\right)\, \text{, } \theta \in \mathbb{R},\label{eq: 1dexample_density}
\end{eqnarray}
with normalising constant $z=\int_{-\infty}^{\infty}q(\theta)d\theta$.
This density has a cusp --- one of the more awkward pathologies of
some naturally occurring densities \citep{bahcall1976star,kato1957eigenfunctions} --- 
and it does not have an analytical integral that easily generalises
to multiple dimensions, but is otherwise a made-up 1-dimensional example that could be interchanged with an other. 

In this instance the Laplace approximation based on the second-order
Taylor expansion at the mode will fail due to the cusp, so we can
use the more robust covariance sampling method. Sampling from the
1d density $q(\theta)$ we find a variance of $\hat{\sigma}^{2}=0.424$,
giving a Gaussian reference density $q_{\text{ref}}(\theta)$ with
normalising constant of $z_{\text{ref}}=1.559$.
The full normalising constant, $z=z_{\text{ref}}\frac{z}{z_{\text{ref}}}$,
is evaluated by Equation \ref{eq: reference_TI}, by setting up a
thermodynamic integration along the sampling path $q^{\lambda}q_{\text{ref}}^{(1-\lambda)}$.
The expectation, $\mathbb{E}_{ q(\lambda;\bm{\theta})}\left[\text{log\,}\frac{q(\bm{\theta})}{q_{\text{ref}}(\bm{\theta})}\right]$,
is evaluated at $5$ points along the coupling parameter
path $\lambda={0.0,0.2,0.5,0.8,1.0}$, shown in Figure \ref{fig: 1dcusp}.
In this simple example, the integral can be easily evaluated to high
accuracy using quadrature \citep{quadpack,scipy}, giving a value of $1.523$. Referenced thermodynamic integration reproduces this value, with convergence of $z$ shown in Figure \ref{fig: 1dcusp}, converging
to $1$\% of $z$ with $500$ iterations and $0.1$\% within $17,000$ iterations.

This example illustrates notable characteristic features
of referenced thermodynamic integration. Here the reference $q_{\text{ref}}(\theta)$
is a good approximation to $q(\theta)$, with $z_{\text{ref}}$
accounting for most ($102$\%) of $z$. Consequently $\frac{z}{z_{\text{ref}}}$
is close to 1, and the expectations, $\mathbb{E}_{ q(\lambda;\bm{\theta})}\left[\text{log\,}\frac{q(\bm{\theta})}{q_{\text{ref}}(\bm{\theta})}\right]$,
evaluated by MCMC for the remaining part of the integral are small. For the same reasons the variance at
each $\lambda$ is small, leading to favourable convergence within a small
number of iterations. And finally $\mathbb{E}_{ q(\lambda;\bm{\theta})}\left[\text{log\,}\frac{q(\bm{\theta})}{q_{\text{ref}}(\bm{\theta})}\right]$
weakly depends on $\lambda$, so there is no need to use a very fine
grid of $\lambda$ values or consider optimal paths---satisfactory
convergence is easily achieved using a simple geometric path with
$4$ $\lambda$ intervals. 

\subsection{2D Pedagogical Example with \emph{Constrained} Parameters}
\begin{figure}
     \centering
    \includegraphics[width=\textwidth]{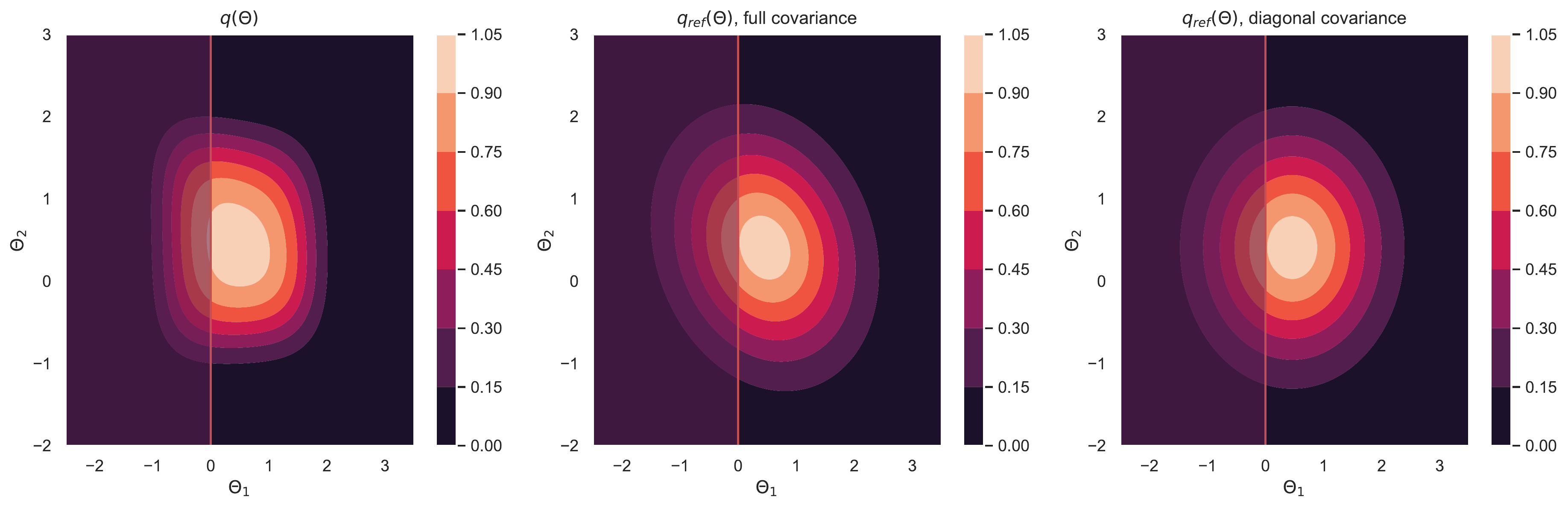}
    \caption{Contour plots of the un-normalised density $q$ and its two reference densities $q_{ref}$, one using a full covariance matrix and another using a diagonal covariance matrix. The red line shows the lower boundary $\theta_1 = 0$ and the shaded $\theta_1 < 0$ region to the left of the line is outside of the support of the density $q$.}
        \label{fig: HessianRefConstrained}
\end{figure}

As a second example, consider a 2-dimensional un-normalised density with a constrained parameter space:
\begin{equation}
    q(\theta_1, \theta_2) = \text{exp}\left[ -\frac{1}{4} \left[ (\theta_1+\frac{1}{2})^2 + (\theta_1+\frac{1}{2})^4 + (\theta_2+\frac{1}{2})^2 + (\theta_2+\frac{1}{2})^4 + \frac{1}{2} \theta_1 \theta_2^2\right]\right]\,,
\end{equation}
where
\[
\theta_1 \in [0,+\infty) \text{ and } \theta_2 \in (-\infty,+\infty)\,.
\]
A reference density $q_\text{ref}(\bm{\theta})$ can be constructed from the Hessian at the mode of $q(\bm{\theta})$.
Notice, that because parameter $\theta_1$ is constrained to be $\geq 0$, integrating the Gaussian approximations $q_\text{ref}(\bm{\theta})$ using the formula given in Equation \ref{eq: reference_integrated}  will give an overestimate. To account for this we use the reference density $q^\text{diag}_\text{ref}(\bm{\theta})$, based on a diagonal Hessian, that has an exact and easy to calculate orthant.
All densities are shown in Figure \ref{fig: HessianRefConstrained}. 

To obtain the log-evidence of the model, we calculated the exact value numerically \citep{scipyNatureMet, quadpack}, using the full covariance Laplace method as per Equation \ref{eq: covariance_laplace} and using the diagonal covariance with correction added to take into account the lower bound of the parameter $\theta_1$, as per Equation \ref{eq: orthant}. The Gaussian reference densities were then used to carry out referenced thermodynamic integration. Results of all methods are given in Table \ref{tab: logEvCorrection}. As expected, without applying the correction the value of the evidence is overestimated.

\begin{table}[ht]
\centering
\begin{tabular}{ll}
\toprule
Method & Evidence \\
\midrule
Exact$^*$ & 3.31\\
Laplace with full covariance & 5.55 \\
Laplace with diagonal covariance + constraint correction & 3.81 \\
Ref TI with full covariance & 4.79 \\
Ref TI with diagonal covariance + constraint correction & 3.33 \\
\bottomrule
\end{tabular}
\caption{Evidence calculated with different methods. Constraint correction refers to imposing the integration limits on the reference as per Equation \ref{eq: orthant}. $^*$ obtained numerically with \citep{scipyNatureMet, quadpack}. \label{tab: logEvCorrection}}
\end{table}

\subsection{Benchmarks---\emph{Radiata Pine}}

To benchmark the application of the referenced TI in the model selection task, two non-nested linear regression models are compared for the \emph{radiata pine} data set \citep{radiata}. This example has been widely used for testing normalising constant calculating methods, since in this instance the exact value of the model evidence can be computed. The data consists of 42 3-dimensional data-points, expressed as $y_i$ - maximum compression strength, $x_i$ - density and $z_i$ - density adjusted for resin content. In this example, we follow the approach of Friel and Wyse \citep{friel_review}, and test which of the two models $M_1$ and $M_2$ provides better predictions for the compression strength:
\[
M_1: y_i = \alpha + \beta(x_i - \bar{x}) + \epsilon_i, \epsilon_i \sim N(0,  \tau^{-1}), i = 1,..., n)\,,
\]
\[
M_2: y_i = \gamma + \delta(z_i - \bar{z}) + \eta_i, \eta_i \sim N(0,  \rho^{-1}), i = 1,..., n)\,.
\]
In other words, we want to know, whether density or density adjusted allows to predict the compression strength better. The priors for the models were selected in a way which enables obtaining an exact solution and can be found in Friel and Wyse \citep{friel_review}.

Five methods of estimating the model evidence were used in this example: Laplace approximation using a sampled covariance matrix, model switch TI along a path directly connecting the models \citep{lartillot, ti_bayesian_vitoratou}, referenced TI, power posteriors with equidistant 11 $\lambda$-placements (labelled here as PP$_{11}$) and power posteriors with 100 $\lambda$-s (PP$_{100}$) as in \citep{friel_review}. For the model switch TI, referenced TI and PP$_{11}$ we used $\lambda \in \{0.0, 0.1, ..., 1.0\}$.

\begin{figure}
    \centering
     \begin{subfigure}[b]{0.49\textwidth}
         \centering
         \includegraphics[width=1\linewidth]{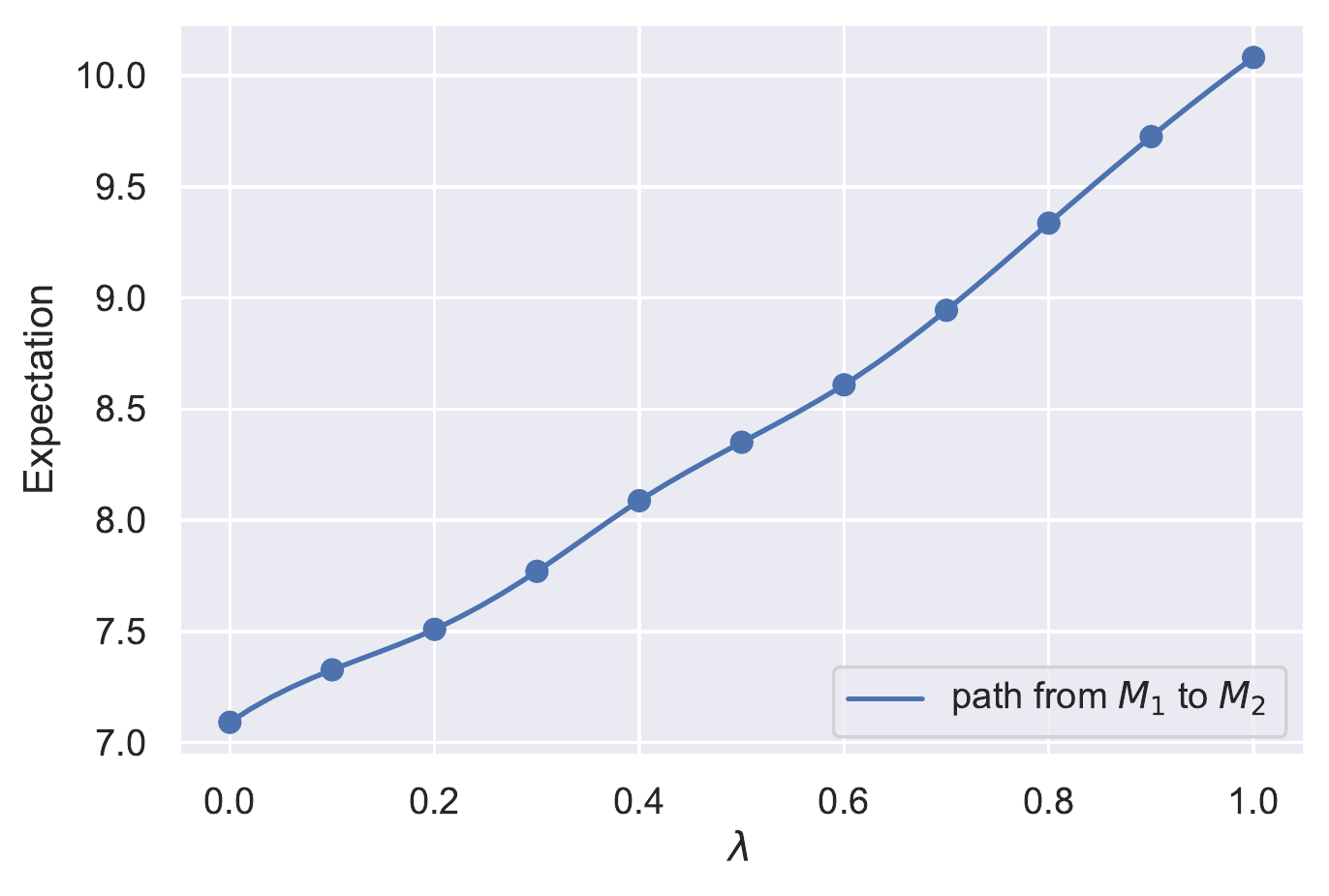}
         \caption{Model switch TI}
     \end{subfigure}
     \hspace{\fill}
     \centering
     \begin{subfigure}[b]{0.49\textwidth}
         \centering
         \includegraphics[width=1\linewidth]{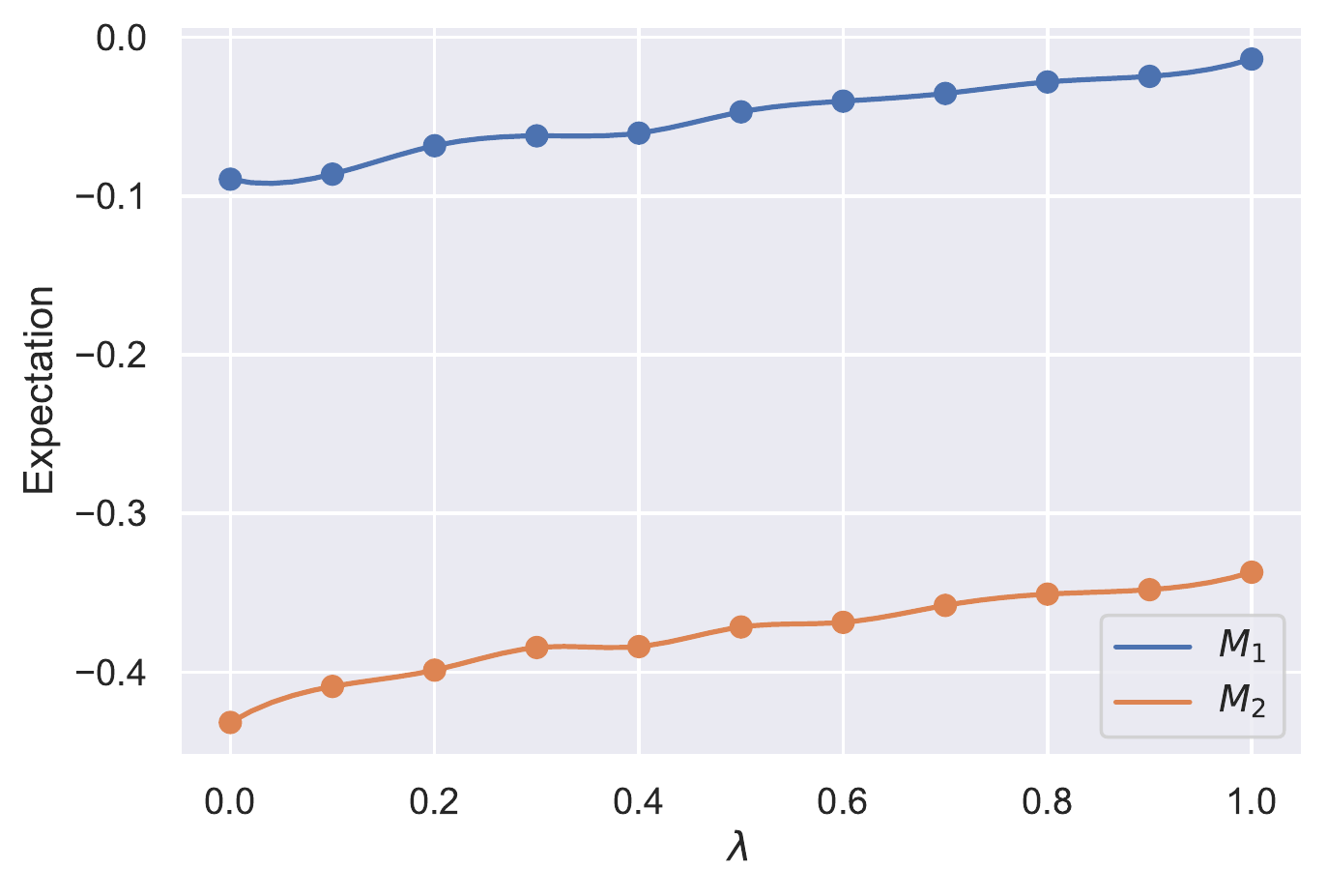}
         \caption{Referenced TI}
     \end{subfigure}
     \hspace{\fill}
     \begin{subfigure}[b]{0.49\textwidth}
         \centering
         \includegraphics[width=1\linewidth]{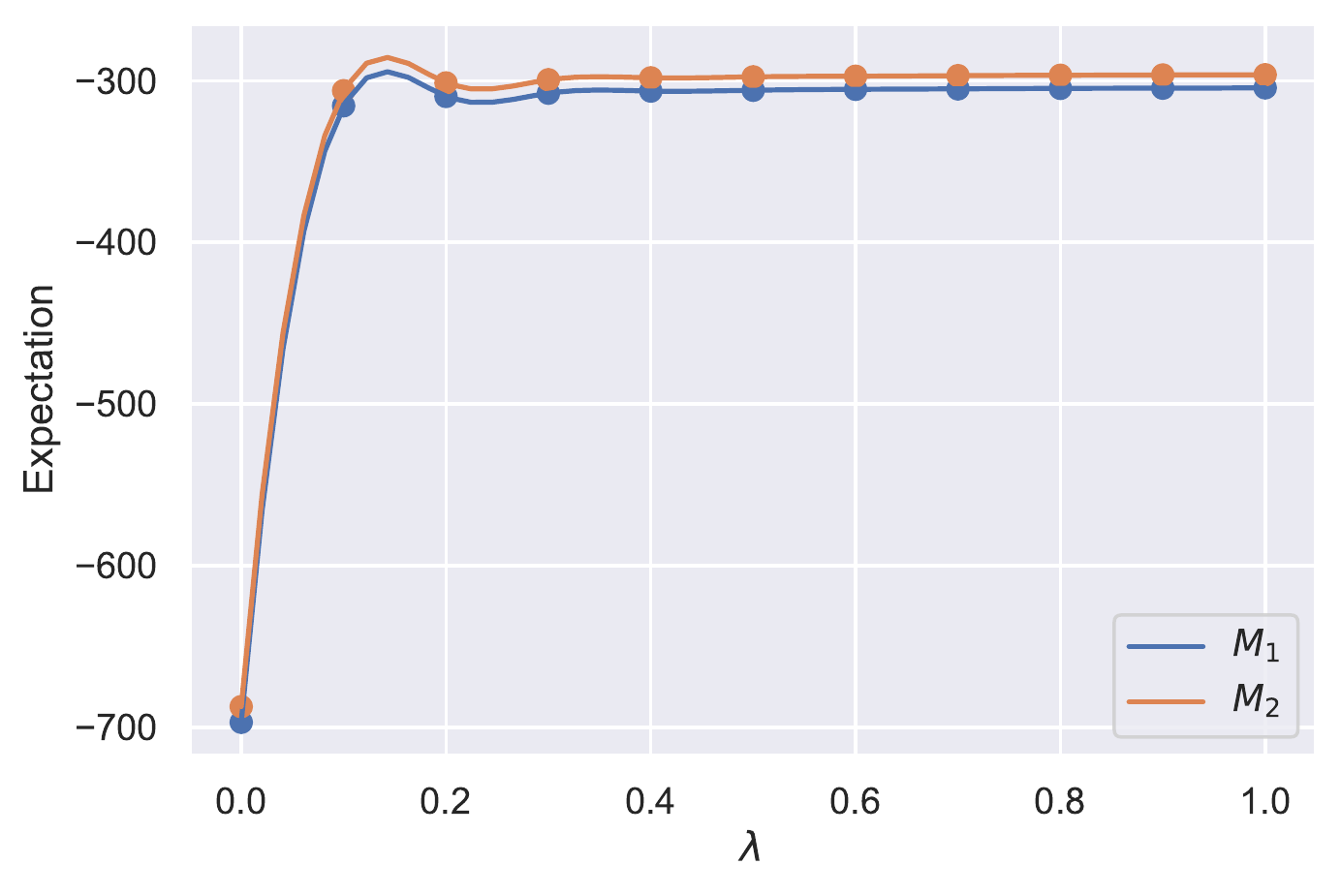}
         \caption{PP$_{11}$}
     \end{subfigure}
    \hspace{\fill}
     \begin{subfigure}[b]{0.49\textwidth}
         \centering
         \includegraphics[width=1\linewidth]{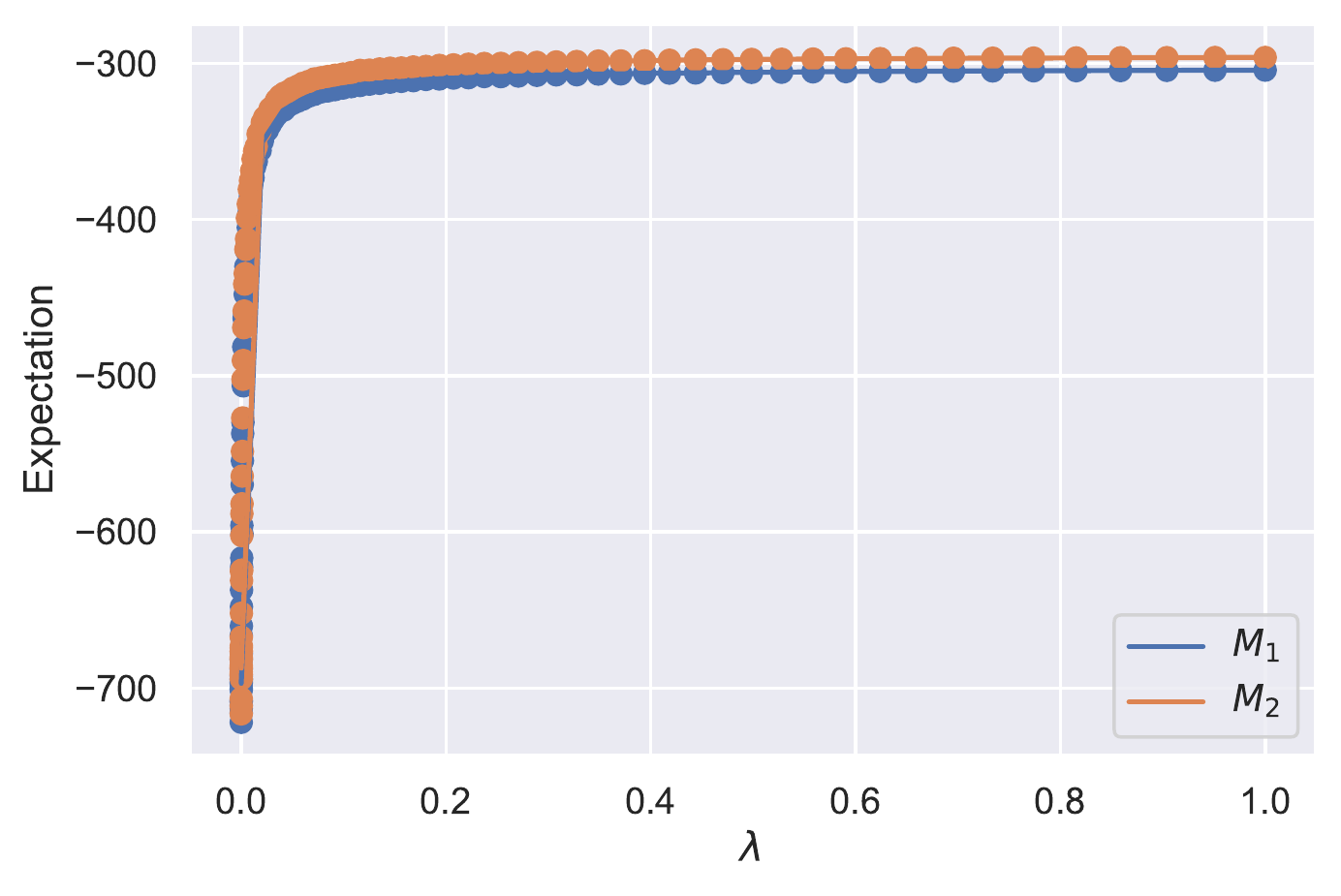}
         \caption{PP$_{100}$}
     \end{subfigure}
        \caption{The HMC-evaluated expectation of $\mathbb{E}_{ q(\lambda;\bm{\theta})}\left[\text{log\,}\frac{q(\bm{\theta})}{q_{\text{ref}}(\bm{\theta})}\right]$ vs coupling parameter $\lambda$ is given for models $M_1$ and $M_2$ for four methods of calculating the model evidence: model switch TI (a), referenced TI (b), power posteriors with 11 $\lambda$-placements (c) and power posteriors with 100 $\lambda$ placements (d). Model switch TI (a) creates the path directly between two competing densities, therefore only one line is shown (see Equation \ref{eq: TI_derivation}).}
        \label{fig: radiata_lambdas}
\end{figure}

The expectation from MCMC sampling per each $\lambda$ for model switch TI, referenced TI, PP$_{11}$ and PP$_{100}$ and fitted cubic splines between the expectations are shown in Figure \ref{fig: radiata_lambdas}. Immediately we notice that both TI methods eliminate the problem of divergence of expectation for $\lambda = 0$, which is observed with the power posteriors, where samples for $\lambda=0$ come from the prior distribution. The PP$_{11}$ method failed to estimate the log-evidence correctly.

\begin{figure}
     \centering
     \begin{subfigure}[b]{0.49\textwidth}
         \centering
         \includegraphics[width=\linewidth]{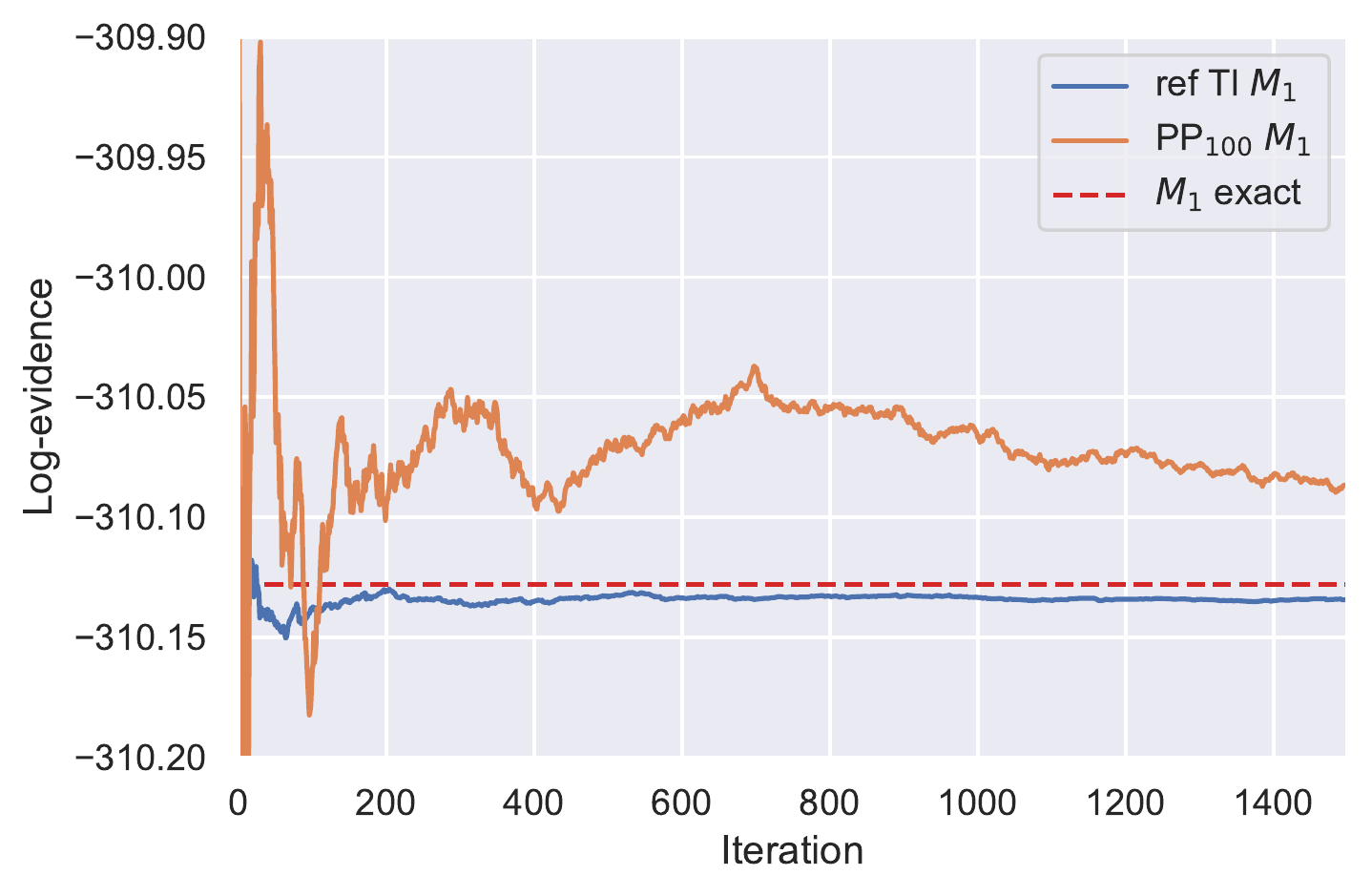}
         \caption{$M_1$}
     \end{subfigure}
     \hspace{\fill}
     \begin{subfigure}[b]{0.49\textwidth}
         \centering
         \includegraphics[width=\linewidth]{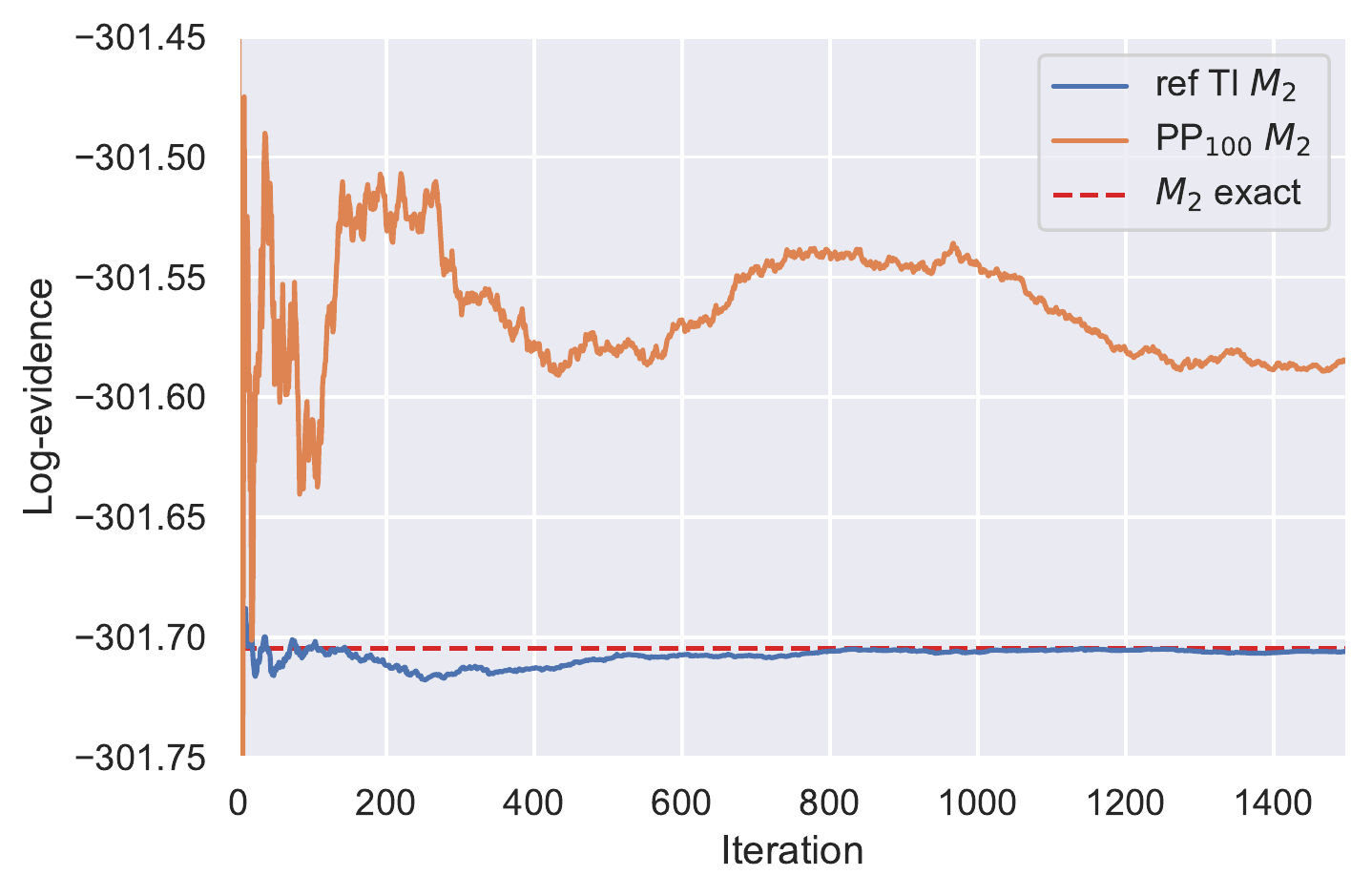}
         \caption{$M_2$}
     \end{subfigure}
    \begin{subfigure}[b]{0.49\textwidth}
         \centering
         \includegraphics[width=\linewidth]{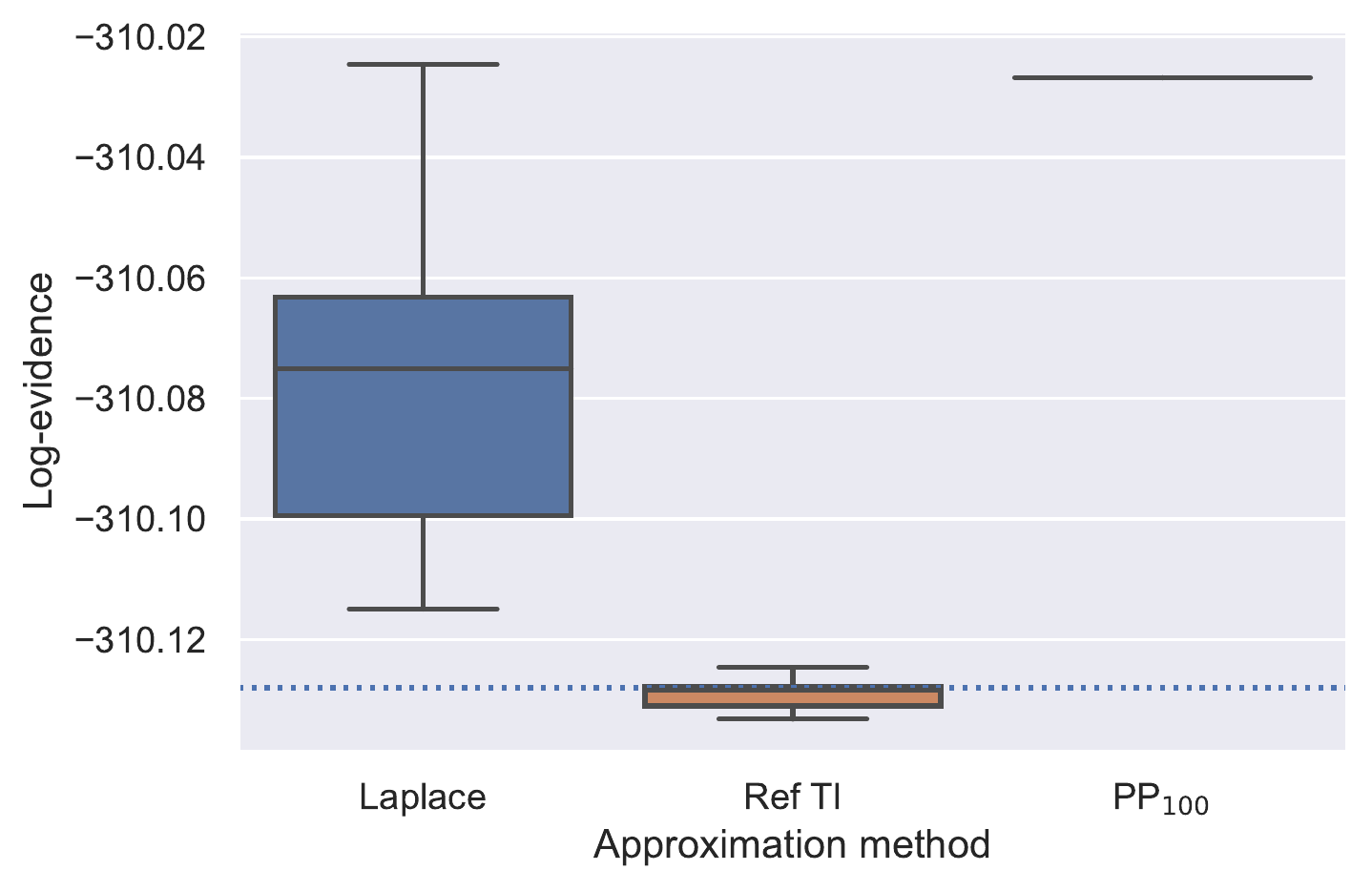}
         \caption{$M_1$}
     \end{subfigure}
     \hspace{\fill}
     \begin{subfigure}[b]{0.49\textwidth}
         \centering
         \includegraphics[width=\linewidth]{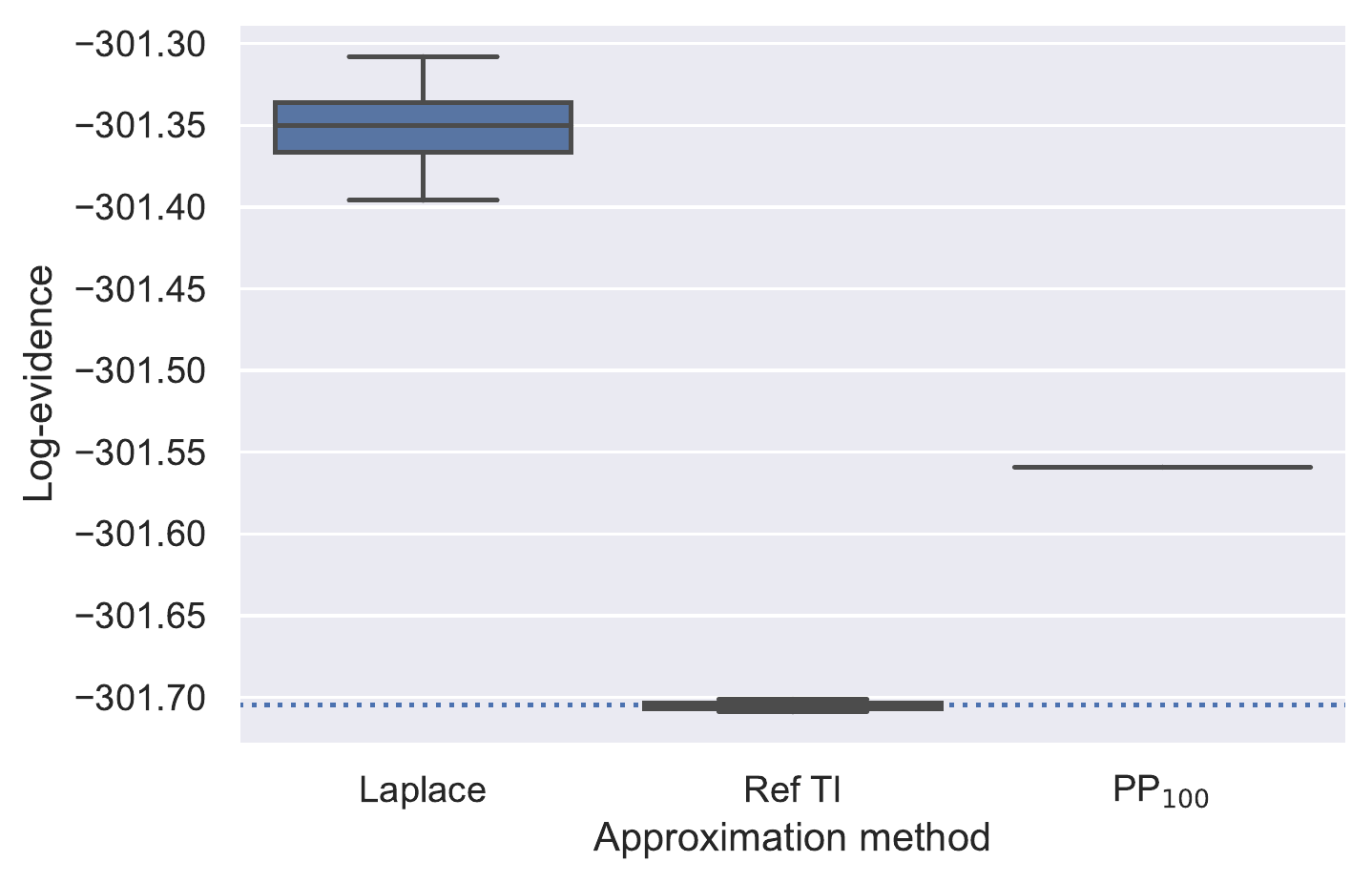}
         \caption{$M_2$}
     \end{subfigure}
        \caption{Log-evidence of $M_1$ and $M_2$ for the three algorithms. (a) and (b) show the rolling mean of log-evidence of $M_1$ and $M_2$ over 1500 iterations per $\lambda$ obtained by referenced TI (blue line) and PP$_{100}$ (orange line) methods. The exact value is shown with red dashed line. (c) and (d) show the mean log-evidence of the two models evaluated over 15 runs of the three algorithms. The exact value of the log-evidence is shown with the dotted line.}
        \label{fig: radiata_evidence}
\end{figure}

The 1-dimensional lines estimated by fitting a cubic spline to the expectation were integrated for each of the models to obtain the log-evidence for $M_1$ and $M_2$, and the log ratio of the two models' evidences for the model switch TI. Rolling mean of the integral over 1500 iterations for referenced TI and PP$_{100}$ are shown in Figure \ref{fig: radiata_evidence} a-b. We can see from the plots, that referenced TI presents excellent convergence to the exact value, whereas PP$_{100}$ oscillates around it. In the same Figure \ref{fig: radiata_evidence}, plots c-d show the distribution of log-evidence for each model generated by 15 runs of the three algorithms: Laplace approximation with sampled covariance matrix, referenced TI and PP$_{100}$. Although all three methods resulted in a log-evidence satisfactorily close to the exact solution, referenced TI was the most accurate and importantly, converged fastest.

The BFs calculated to assess whether model $M_2$ fits the data better than model $M_1$ and the number of iterations needed to achieve standard error of 0.5\%, excluding the iterations needed for the MCMC burn-in, are presented in Table \ref{tab: bf_pines}. Notably, both TI methods gave a BF very close to the exact value. Referenced TI performed the best out of tested methods---it converged faster than all other methods, requiring only 308 MCMC draws compared to 55,000 draws needed for the power posterior method or over 2,000 for the model switch TI. Referenced TI also showed excellent accuracy both in estimating individual model's evidence and the BF.

\begin{table}[!h]
\centering

\begin{tabular}{lcc}
\toprule
Method & $BF_{21}$ & MCMC steps\\
\midrule
Exact & 4552.35 & - \\
Laplace approximation$^*$ & 6309.10 & -  \\
Model switch TI  & 4557.63 & 2,365 \\ 
Referenced TI & 4558.71 & 308 \\ 
PP$_{11}$ & 4463.71 & 41,514 \\ 
PP$_{100}$ & 4757.82 & 55,000 \\ 
\bottomrule
\end{tabular}
\caption{Comparison of Bayes factors for \emph{radiata pine} models for each method. Here we show $BF_{21} = \frac{M_2}{M_1}$ to determine whether model $M_2$ is better than model $M_1$. Both TI and referenced TI methods used 11 equidistant $\lambda$-s. Power posteriors method was used with 11 (PP$_{11}$) and 100 (PP$_{100}$) $\lambda$-s. Third column shows the total number of MCMC steps required to achieve standard error of 0.5\%, excluding the burn-in steps. $^*$ - using sampled covariance matrix. \label{tab: bf_pines}}
\end{table}

\subsection{Model Selection for the COVID-19 Epidemic in South Korea}
The final example of using referenced TI for calculating model evidence is fitting a renewal model to COVID-19 case data from South Korea. The data were obtained from  \url{https://opendata.ecdc.europa.eu/covid19/casedistribution/csv} (accessed 19-07-2020) and contained time-series data of confirmed cases from 31-12-2019 to 18-07-2020. The model is based on the Bellman-Harris branching process whose expectation is the renewal equation. Its derivation and details are explained in Mishra et al. \citep{mishra2020derivation} and a short explanation of the model is provided in the Appendix. Briefly, the model is fitted to the time-series case data and estimates a number of parameters, including serial interval and the effective reproduction number, $R_t$. Number of cases for each day are modelled by a negative binomial distribution, with shape and overdispersion parameters estimated by a renewal equation. Three modification of the original model were tested:
\begin{itemize}
  \item $GI=k$, $k = 5,6,6.5,7,8,9,10,20$---fixing the mean of the $GI$ parameter, denoting the mean of Rayleigh-distributed generation interval (which we assume to be the same as the serial interval),
  \item $AR(k)$, $k = 2,3,4$---autoregressive model with $k$ days lag,
  \item $W=k$, $k = 1,2,3,4,7$---changing the length $k$ of the sliding window $W$.
\end{itemize}

Within each group of models, $GI$, $AR$ and $W$, we want to select the best model through the highest evidence method. For example, we want to check whether $GI=6$ fits the data better than $GI=10$, etc. The dimension of each model was dependent on the modifications applied, but in all the cases the normalising constant was a 40- to 200-dimensional integral. The log-evidence of each model was calculated using the Laplace approximation with a sampled covariance matrix, and then correction to the estimate was obtained using referenced TI method. Values of the log-evidence for each model calculated by both Laplace and referenced TI methods are given in Table \ref{tab: LogEvidenceBH}. Interestingly, the favoured model in each group, that is the model with the highest log-evidence, was different when the evidence was evaluated using the Laplace approximation than when it was evaluated with referenced TI. For example, using the Laplace method, sliding window of length 7 was incorrectly identified as the best model, whereas with referenced TI window of length 2 was chosen to be the best among the tested sliding windows models, which agrees with the previous studies of the window-length selection in H1N1 influenza and SARS outbreaks \citep{informationTheory_Parag_Christl}. \emph{This exposes how essential it is to accurately determine the evidence, even good approximations can result in misleading results}. Log-Bayes factors for all model pairs within each of the three groups are shown in Figure \ref{fig: AppendixBayes_factors_BH} in the Appendix.

\begin{table}[ht]
    \centering
    \begin{tabular}{@{}cccc@{}}
        \toprule
        Model & Log-evidence (Laplace) & Correction & Log-evidence (ref TI) \\
        \midrule
        GI=5 & -1274 & 558 & -716 [-715.6, -715.2] \\
        GI=6 & -1274 & 572 & -703 [-703.3, -702.7] \\
        GI={6.5} & -1269 & 530 & -739 [-738.6, -738.3] \\
        GI=7 & -1255 & 522 & -732 [-732.4,  -731.8] \\
        GI=8 & -1245 & 561 & \textbf{-685} [-685.5, -684.7] \\
        GI=9 & -1310 & 507 & -803 [-802.8, -802.3]  \\
        GI=10 & -1313 & 508 & -805 [-805.1, -805.3] \\
        GI=20 & \textbf{-1170} & 385 & -796 [-796.3, -795.5] \\
        \hline
        AR(2) & \textbf{-1207} & 496 & -711 [-711.2, -710.6] \\
        AR(3) & -1293 & 589 & \textbf{-704} [-704.7, -703.7] \\
        AR(4) & -2166 & 1346 & -821 [-820.6, -819.2.] \\
        \hline
        W=1 & -1260 & 458 & -802 [-802.1, -801.6] \\
        W=2 & -1069 & 278 & \textbf{-791}  [-791.2, -790.7] \\
        W=3 & -1003 & 196 & -807 [-807.5, -807.2] \\
        W=4 & -940 & 129 & -811 [-811.1, -810.7] \\
        W=7 & \textbf{-875} & 62 & -814 [-813.7, -813.5] \\
        \bottomrule
    \end{tabular}
\caption{Log-evidence estimated by Laplace approximation, added referenced TI correction and total log-evidence from referenced TI, with 95\% credible interval given in brackets. In each section, model with the highest log-evidence estimated by Laplace or referenced TI method is indicated in bold. \label{tab: LogEvidenceBH}}
\end{table}

\subsection{Interpretation of the COVID-19 model selection}

\begin{figure}
     \centering
     \begin{subfigure}[b]{1\textwidth}
         \centering
         \includegraphics[scale = 0.5]{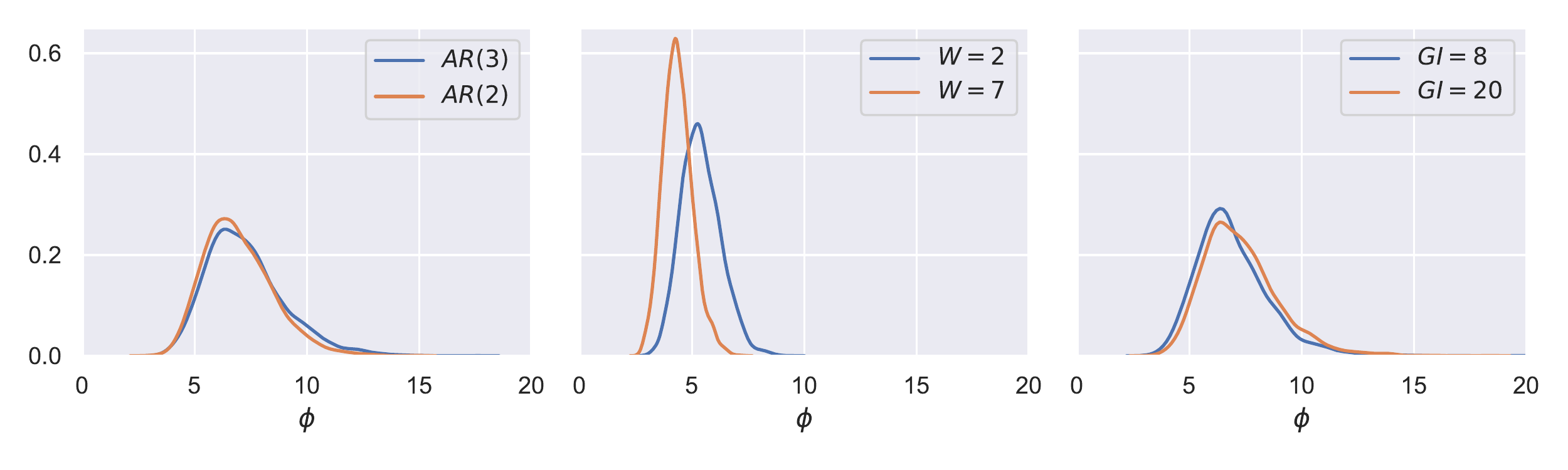}
         \caption{Posterior distributions for overdispersion parameter $\phi$}
     \end{subfigure}
     \hspace{\fill}
     \begin{subfigure}[b]{1\textwidth}
         \centering
         \includegraphics[scale = 0.5]{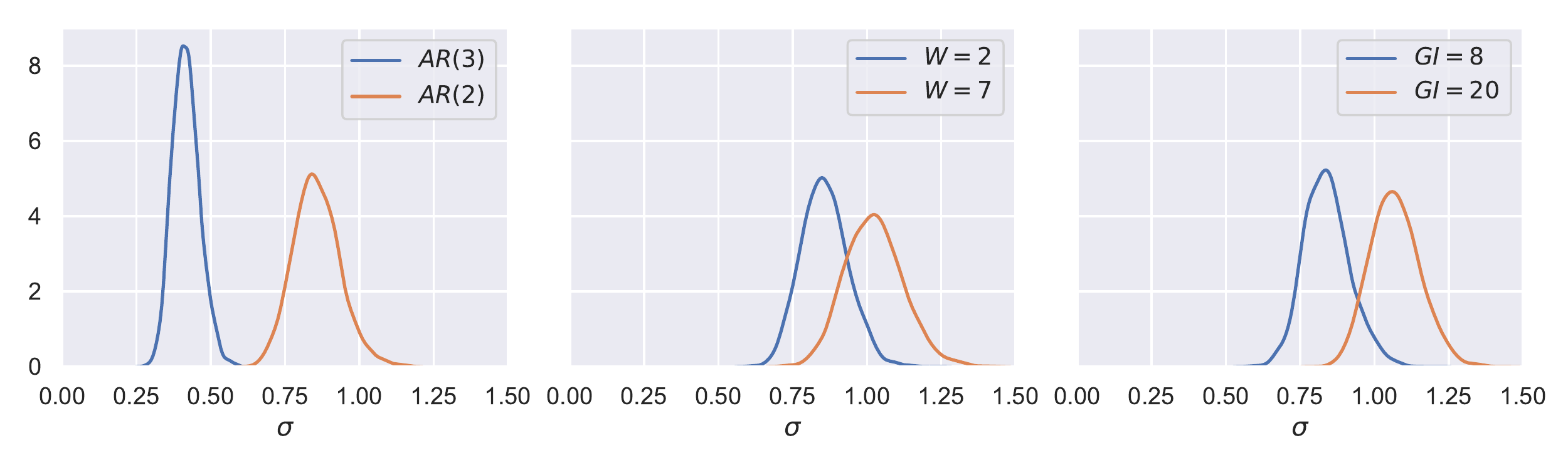}
         \caption{Posterior distributions for $\sigma$ parameter}
     \end{subfigure}
    \hspace{\fill}
     \begin{subfigure}[b]{1\textwidth}
         \centering
         \includegraphics[scale = 0.5]{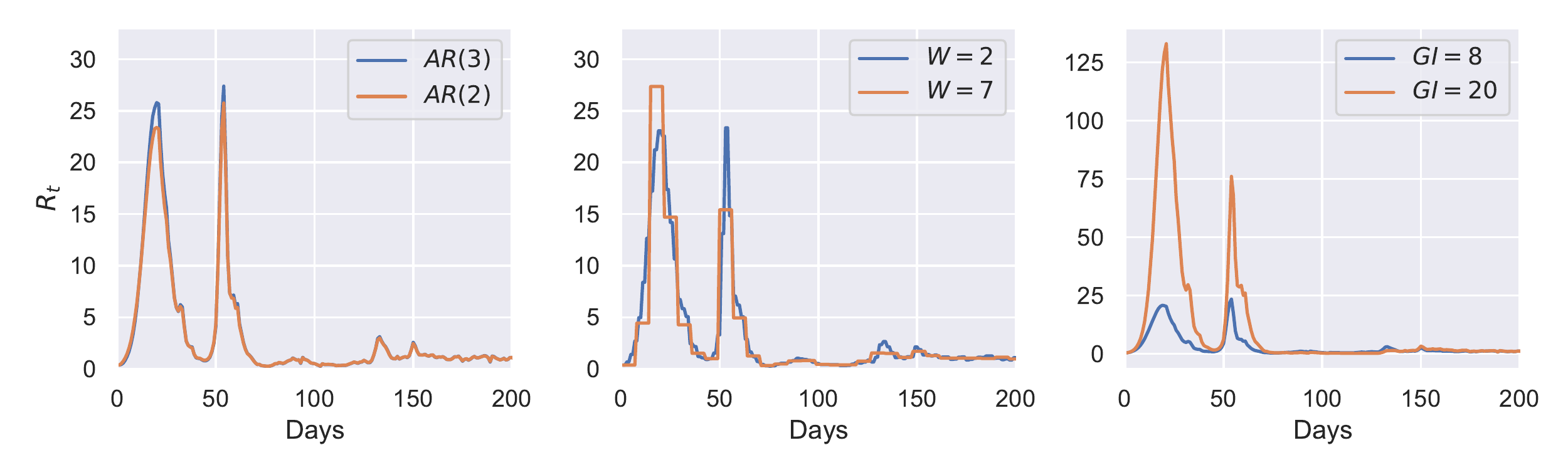}
         \caption{$R_t$ generated by the favoured models}
     \end{subfigure}
        \caption{Posterior distributions for models' parameters for models favoured by BFs using the Laplace approximation (orange lines) and referenced TI (blue lines). \label{fig: bh_posteriors}}
\end{figure}

The importance of performing model selection in a rigorous way is clear from Figure \ref{fig: bh_posteriors}, where the posterior densities of parameters $\phi$ and $\sigma$ and the generated $R_t$ time-series are plotted for the models favoured by Laplace and referenced TI methods (meaning of the parameters is given in the Appendix). The differences in the densities and time-series show the pitfalls of selecting an incorrect model. For example, the parameter $\sigma$ was overestimated by the models selected by Laplace approximation in comparison to these selected by the referenced TI. The differences between the two favoured models were most extreme for the $GI=8$ and $GI=20$ models. While a $GI=8$ is plausible, even likely for COVID-19, $GI=20$ is implausible given observed data \citep{Bi_covid_serial_interval}. This is further supported by observing that for $GI=20$, favoured by the Laplace method, $R_t$ reached the value of over 125 in the first peak---around 100 more than for the $GI=8$. The second peak was also largely overestimated, where $R_t$ reached a value of 75.

We find it interesting to note that all models present a similar fit to the confirmed COVID-19 cases data, as shown by Figure \ref{fig: appendixCasesFit} in the Appendix. This makes it impossible to select the best model through visual inspection and comparison of the model fits, or by using model selection methods that do not take the full posterior distributions into account. Although the models might fit the data well, other quantities generated, which are often of interest to the modeller, might be completely incorrect. Moreover, it emphasises the need to test multiple models before any conclusion or inference is undertaken, especially with the complex, hierarchical models. In epidemiology this is important as the modellers can be tempted to pick arbitrary parameters for their model, as long as the predictions fit the data. Although the fit might be accurate, other inferred parameters or uncertainty related to the predictions might be completely inappropriate for making any meaningful predictions.

In the \emph{radiata pine} example, the contribution of TI to the marginalised likelihood estimated by the Laplace approximation was not substantial and using the Laplace approximation would suffice to make an informed model choice. However in this example, we see that the TI contribution is relatively large, and that it changes the decision to be made based on model evidence relative to the Laplace method. Moreover, from Table \ref{tab: LogEvidenceBH} we see that the evidence was the highest for the "boundary" models when Laplace approximation was applied. For example, for the sliding window length models, when the Gaussian approximation was applied, the log-evidence was monotonically increasing with the value of $W$ within the range of values that seem reasonable ($W = 1-7$). In contrast, with referenced TI, the log-evidence is concave within the range of \emph{a priori} reasonable parameters.

\section{Discussion \label{sec: Discussion}}

The examples shown in Section \ref{sec: Application} illustrate the applicability of the referenced TI algorithm for calculating model evidence. In the \emph{radiata pine} example, referenced TI performed better than the other tested methods in terms of accuracy and speed. The power posteriors method required a much denser placement of the coupling parameters around $\lambda = 0$, where the values are sampled purely from the prior distribution. In the case of referenced TI, at $\lambda = 0$ values are sampled from the reference density, which should be closer to the original density (in the sense of Kullback–Leibler or Jensen-Shannon divergence), which results not only in a more accurate estimate of the normalising constant, but also much faster convergence of the MCMC samples. It also worth noting that referenced TI even performed better than the model switch TI method. A detailed theoretical characterisation of rates of convergence is beyond the scope of this article, nonetheless the empirical tests presented have consistently shown faster convergence than with comparative approaches. This is useful to know in the context of evaluating model evidence in complex hierarchical models where where each MCMC iteration is computationally demanding. 

\begin{figure}
\begin{centering}
\includegraphics[scale=1.5]{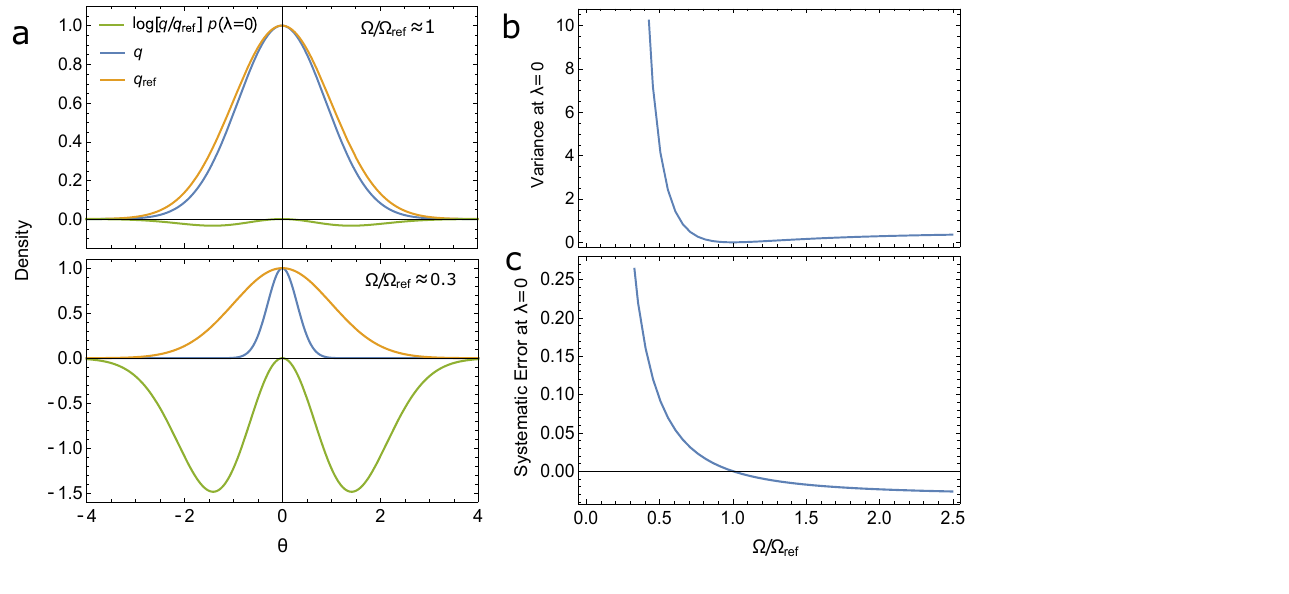}
\par\end{centering}
\caption{a) 1D examples to illustrate the bias and variance introduced with finite MCMC samples when $q$ and $q_\text{ref}$ are mismatched. In these examples $\Omega$ and $\Omega_\text{ref}$ denote the domain of the $99\%$ quartiles of $q$ and $q_\text{ref}$. 
b) A mismatch between $q$ and $q_\text{ref}$ ($\Omega$ and $\Omega_\text{ref}$) causes the variance of $\text{log} \frac{q}{q_\text{ref}}$ to increase, requiring more iterations to convergence.
c) Similarly the mismatch causes the mass of the distribution for the expectation of $\text{log} \frac{q}{q_\text{ref}}$ (evaluated with respect to the reference distribution) to increase beyond the parameter range effectively sampled with finite iterations, in this example corresponding to the $99\%$ quartile of the sampling distribution, thus introducing a bias in the expectation.} \label{fig: match_density}
\end{figure}

Although referenced thermodynamic integration and other methods using path-sampling have theoretical asymptotically exact Monte Carlo estimator limits, in practice a number of considerations affect accuracy. For example, biases will be introduced to the referenced TI estimate in practice if one endpoint density substantially differs from another. Then the volume of parameter space that must be explored to produce an unbiased estimate of the expectation cannot be sampled based on the reference density generating proposals within a practical number of iterations. The point is shown for a simple 1D example in Figure \ref{fig: match_density}. Similarly, the larger the mismatch, the higher the variance and slower the expectation is to converge. 
This illustrates the advantage of using
a reference that matches the posterior as closely as possible, as
opposed to a typically wide reference like the prior distribution, that gives
the characteristic divergence at $\lambda=0$ with power posteriors. Measures of density similarity in path sampling have been discussed by \citep{lefebvre2010path}, however in practical terms there remains much scope for analysis of reference performance in terms of scaling with distribution dimension and type, which should be considered in detail in future work.

Furthermore, the discretisation of the coupling parameter path in $\lambda$ can introduce a discretisation bias. For the power posteriors method, Friel et al. (2017) propose an iterative way of selecting the $\lambda$-placements to reduce the discretisation error \citep{friel_improving}. Calderhead and Girolami (2009) test multiple $\lambda$-placements for 2 and 20 dimensional regression models, and report relative bias for each tested scenario \citep{calderhead_temperatures}. In the referenced TI algorithm discretisation bias is however negligible --- the use of the reference density results in TI expectations that are both small and have low variance, and therefore curvature with respect to $\lambda$. In our framework we use geometric paths with equidistant coupling parameters $\lambda$ between the un-normalised posterior densities, but there are other possible choices of the path constructions, for example a harmonic \citep{gelman1998simulating} or hypergeometric path \citep{ti_bayesian_vitoratou}. This optimisation might be worthwhile exploring, however, as illustrated in Fig. 3b, the expectations evaluated vs $\lambda$ are typically near-linear with referenced TI suggesting limited gains, although the extent of this will differ from problem to problem.

In the application to the renewal model for the COVID-19 epidemic in South Korea, we showed that for a complex structured model, hypothesis selection by Laplace approximation of the normalising constant can give misleading results. Using referenced TI, we calculated model evidence for 16 models, which enabled a quick comparison between chosen pairs of competing models. Importantly, the evidence given by the referenced TI was not monotonic with the increase of one of the parameters, which was the case for the Laplace approximation in the models tested. The referenced TI presented here will similarly be useful in other situations particularly where the high-dimensional posterior distribution is uni-modal but non-Gaussian.

\section{Conclusions \label{sec:Conclusions}}
Normalising constants are fundamental in Bayesian statistics and allow the best model to be selected for given data. In this paper we give an account of referenced thermodynamic integration (TI), in terms of theoretical consideration regarding the choice of reference, and show how it can be applied to realistic practical problems. We show referenced TI allows efficient calculation of a single model's evidence by sampling from geometric paths between the un-normalised density of the model and a judiciously chosen reference density --- here, a sampled multivariate normal that can be generated and integrated with ease. The referenced TI approach was applied to several examples, in which normalising constants over 1 to 200 dimensional integrals were calculated. In the examples, the referenced TI approach had better convergence performance in terms of iterations to a cut-off convergence and minimum bias achievable compared to similar methods such as power posteriors or model switch thermodynamic integration.
We showed the referenced TI method has practical utility for substantially challenging problems of model selection --- in this instance concerning the epidemiology of infectious diseases --- and suggest similar applicability in other fields of applied machine learning that rely on high-dimensional Bayesian models with non-inferential hyper-parameters.

\cleardoublepage{}

\printbibliography

\cleardoublepage{}

\appendix
\section*{Appendix A. COVID-19 Model\label{sec:Appendix}}


The COVID-19 model shown is based on the renewal equation derived from the Bellman-Harris process. The details of the model and its derivation are provided in Mishra et al. \citep{mishra2020derivation}. Here, we give a short overview of the $AR(2)$ model.
The model has a Bayesian hierarchical structure and is fitted to the time-series data containing a number of new confirmed COVID-19 cases per day in South Korea, obtained from \url{https://opendata.ecdc.europa.eu/covid19/casedistribution/csv}. New infections $y(t)$ are modelled by a negative binomial distribution, with a mean parameter in a form of a renewal equation. The number of confirmed cases $y(t)$ is modelled as:
\[
y \sim \text{NegBin}(f(t), \phi)\,,\\
\]
where $\phi$ is an overdispersion or variance parameter and the mean of the negative binomial distribution is denoted as $f(t)$ and represents the daily case data through:  
\[
f(t) = R_0 \int^t_{\tau=0} f(t-\tau)g(\tau)d\tau\,.
\]
As the case data is not continuous but is reported per day, $f(t)$ can be represented in a discretised, binned form as:
\[
f(t) = R_t \sum _{\tau < t} f(t-\tau)g(\tau)\,.
\]
Here, $g(\tau)$ is a Raleigh-distributed serial interval with mean $GI$, which is discretised as 
\[g_s = \int ^{s + 0.5}_{s-0.5} g(\tau)d\tau \text{ for } s=2,3,... \text{ and } g_1 = \int ^{1.5}_{0} g(\tau)d\tau\,.
\]
$R_t$, the effective reproduction number, is parametrised as $R_t = \text{exp}(\epsilon_t)$, with exponent ensuring positivity. $\epsilon_t$ is an autoregressive process with two-days lag, that is AR(2), with $\epsilon_1 \sim N(-1, 0.1)$, $\epsilon_2 \sim N(-1, \sigma)$ and
\[
\epsilon_t \sim N(\rho_1\epsilon_{t-1} + \rho_2\epsilon_{t-2}, \sigma_t) \text{ for } t=\{3,4,5,...\}.
\]

The model's priors are:
\begin{align*} 
\sigma &\sim N^+(0,0.2)\,,\\
\rho_{1} &\sim N^+(0.8,0.05)\,, \\
\rho_{2} &\sim N^+(0.1,0.05)\,, \\
\phi &\sim N^+(0,5)\,, \\ 
GI &\sim N^+(0.01,001)\,. \\ 
\end{align*}

Modification were applied to this basic model, to obtain the different variants of the model as described in Section \ref{sec: Application}. First group of models analysed was the AR(2) model described above, but with the $GI$ parameter fixed to a certain value instead of inferring that parameter from the data. $AR(3)$ and $AR(4)$ models had additional parameters $\rho_3$ and $\rho_4$, which allow to model the autoregressive process with a longer lag (3- and 4- days respectively). Finally, models $W=k$, $k=1,..,7$ were similar to the $AR(2)$ model, but the underlying assumption of these models is that the $R_t$ stays constant for the duration of the length of the sliding window $W=k$.


\begin{figure}
     \centering
     \begin{subfigure}[b]{0.7\textwidth}
         \centering
         \includegraphics[width=1\linewidth]{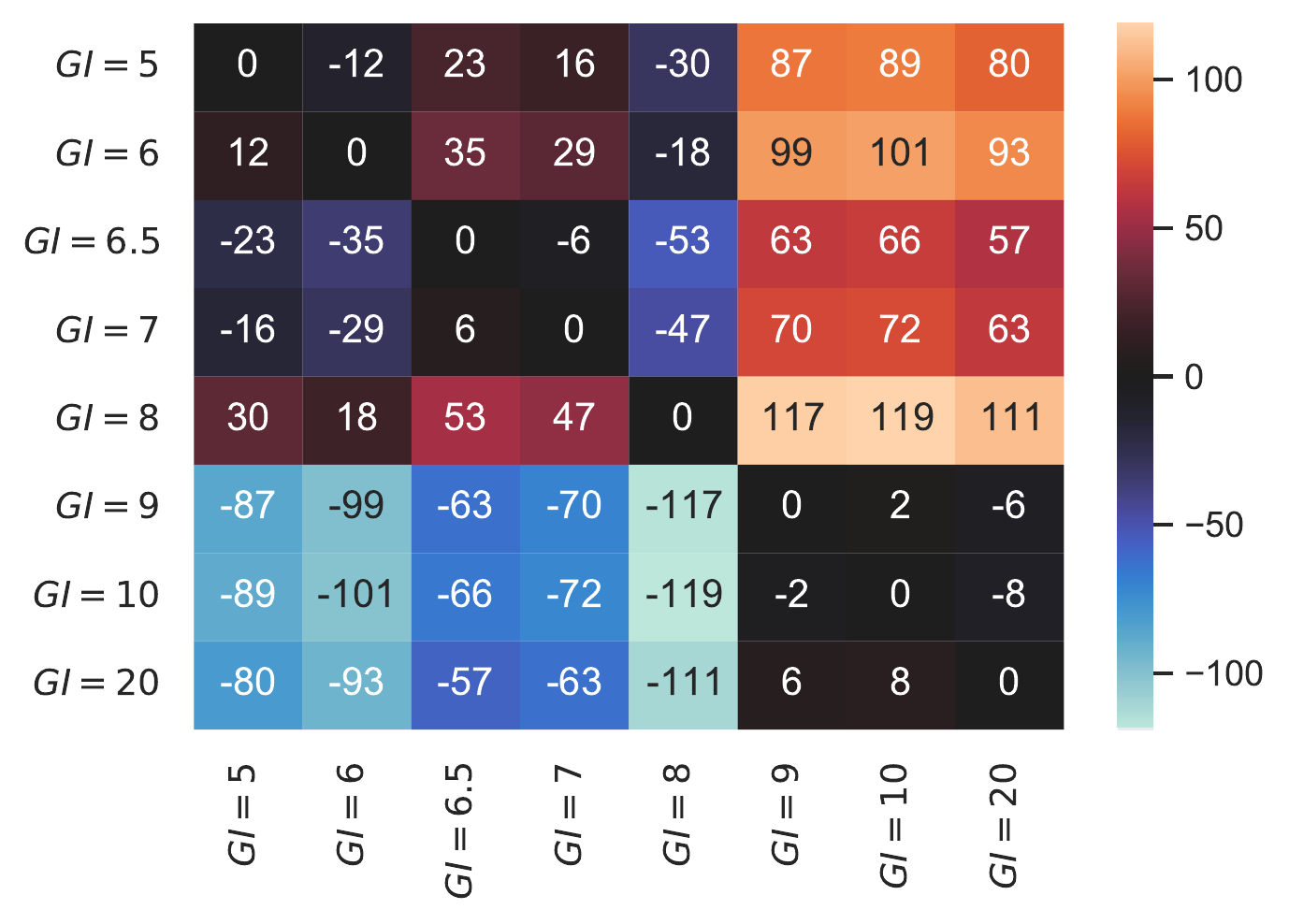}
         \caption{}
     \end{subfigure}
     \hspace{\fill}
     \begin{subfigure}[b]{0.49\textwidth}
         \centering
         \includegraphics[width=1\linewidth]{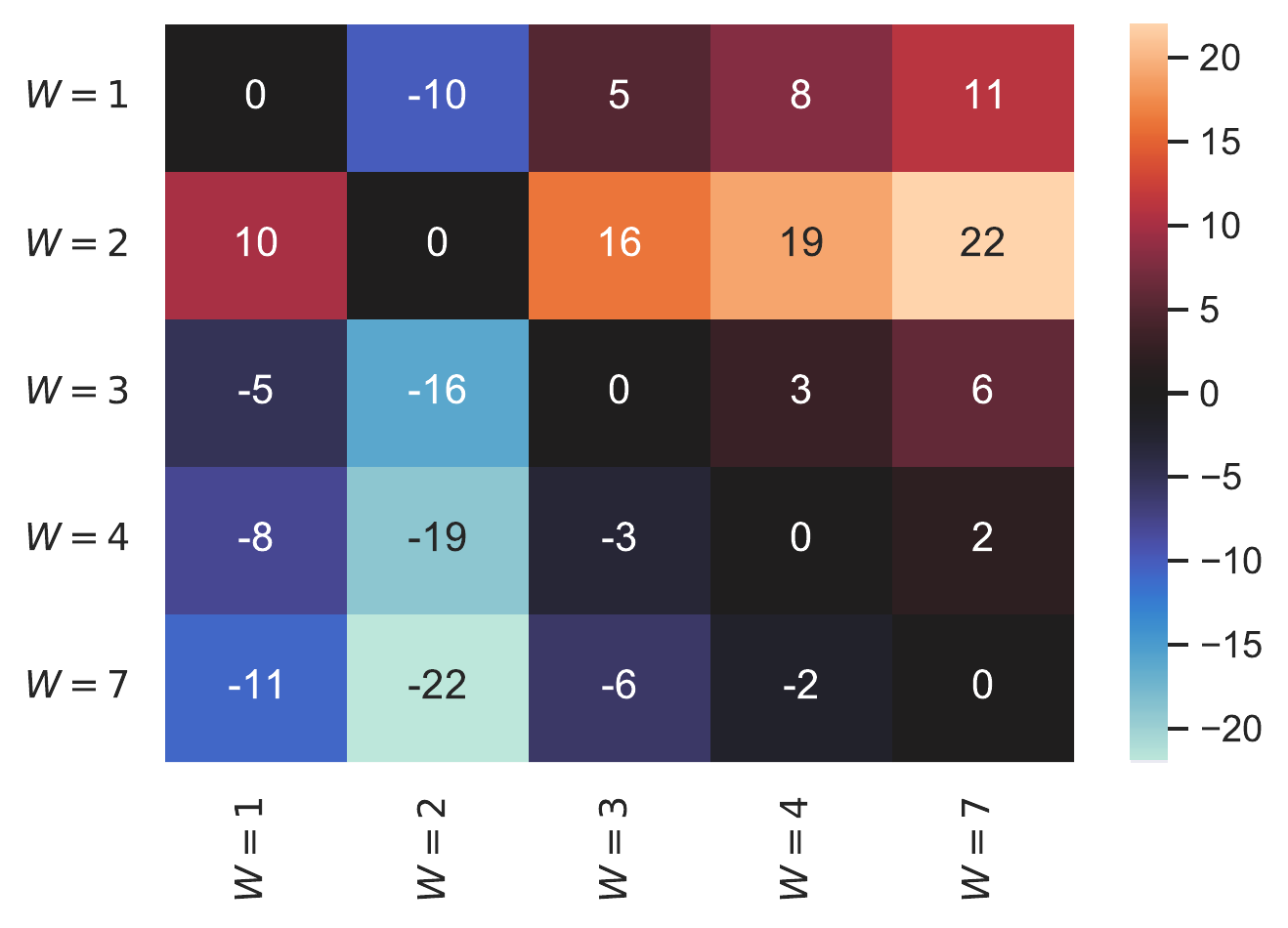}
         \caption{}
     \end{subfigure}
    \hspace{\fill}
     \begin{subfigure}[b]{0.49\textwidth}
         \centering
         \includegraphics[width=1\linewidth]{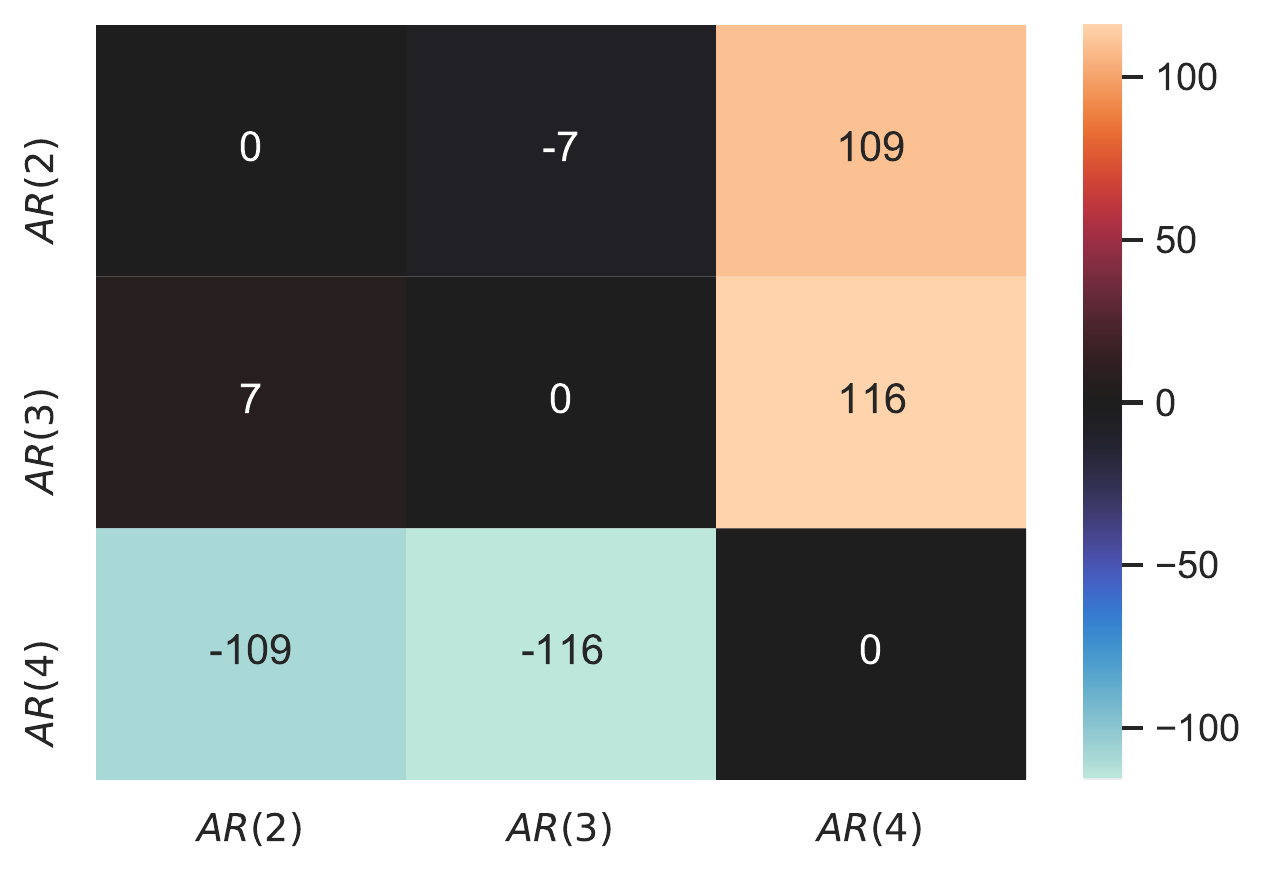}
         \caption{}
     \end{subfigure}
        \caption{Logarithms of Bayes factors for the analysed COVID-19 renewal models, evaluated using the normalising constants ratios obtained by referenced TI. In each cell, the colour indicates the value of the $BF_{1,2}$ for models $M_1$ (row) and $M_2$ (column). Higher values, that is a brighter orange colour, suggest that $M_1$ is strongly better than $M_2$, and values below 0 in blue palette indicate that $M_1$ is worse than $M_2$. $GI=8$ performed best out of fixed $GI$ models, $W=2$ best out of sliding window models, and $AR(3)$ performed better than $AR(2)$ and $AR(4)$. For the interpretation of the BF values see \citep{bayes_factors}.
        \label{fig: AppendixBayes_factors_BH}}
\end{figure}

\begin{figure}
     \centering
    \includegraphics[width=\textwidth]{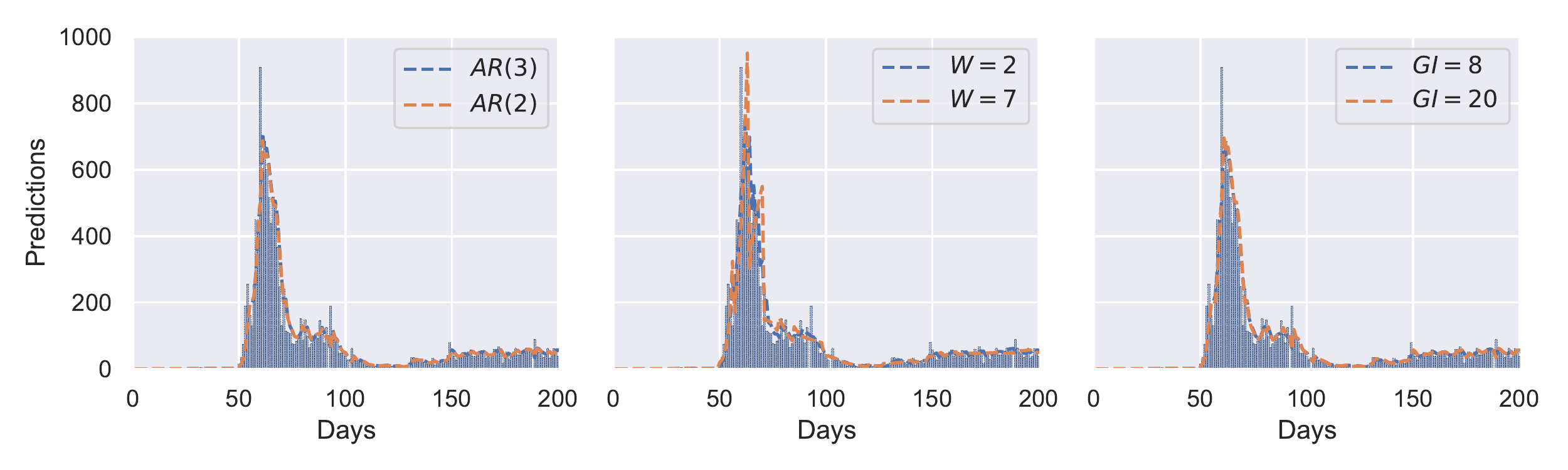}
    \caption{Cases of SARS-CoV-2 infections in South Korea from the data (shown with bars) and the cases predicted by different models. On each graph, predictions made by the model favoured by the Laplace approximation is shown with a blue dashed line, and predictions made by the referenced TI favoured models are shown with an orange dashed line. The lines in all three subplots are largely overlapping, revealing that all models fitted the case data similarly well.}
        \label{fig: appendixCasesFit}
\end{figure}

\end{document}